\documentclass[bm,aps,%showpacs,
amsfonts,amssymb,
preprint,
nofootinbib]{revtex4}

\usepackage[dvipdfmx]{graphicx}
\usepackage{natbib}
\usepackage{amsmath}

\usepackage{color}
\font\mybb=msbm10 at 12pt

\font\mybbsub=msbm10 at 8pt
\font\mybbsmall=msbm10 at 10pt

\def\bb#1{\hbox{\mybb#1}}

\def\bbsub#1{\hbox{\mybbsub#1}}
\def\bbsmall#1{\hbox{\mybbsmall#1}}

\def\ZZ {\bb{Z}}

\def\ZZsmall {\bbsmall{Z}}

\def\PP {\bb{P}}
\def\PPsub {\bbsub{P}}
\def\PPsmall {\bbsmall{P}}

\def\AA{{\bf A}}
\def\BB{{\bf B}}
\def\CC{{\bf C}}

\newcommand\beqa{\begin{eqnarray}}
\newcommand\eeqa{\end{eqnarray}}
\newcommand\n{\nonumber\\}

\begin{document}

{~}

\title{
A dessin on the base: a description of mutually non-local 7-branes without using branch cuts
}
\vspace{2cm}
\author{Shin Fukuchi\footnote[1]{E-mail:fshin@post.kek.jp},
Naoto Kan\footnote[2]{E-mail:naotok@post.kek.jp},
Shun'ya Mizoguchi\footnote[3]{E-mail:mizoguch@post.kek.jp}
and Hitomi Tashiro\footnote[4]{E-mail:tashiro@post.kek.jp}}

%\vspace{1cm}

\affiliation{\footnotemark[3]Theory Center, 
Institute of Particle and Nuclear Studies,
KEK\\Tsukuba, Ibaraki, 305-0801, Japan 
}

\affiliation{\footnotemark[1]\footnotemark[2]\footnotemark[3]\footnotemark[4]SOKENDAI (The Graduate University for Advanced Studies)\\
Tsukuba, Ibaraki, 305-0801, Japan 
}

\begin{abstract} 
We consider the special roles of the zero loci of 
the Weierstrass invariants $g_2(\tau(z))$, $g_3(\tau(z))$  
in F-theory on an elliptic fibration over $\PPsmall^1$ 
or a further fibration thereof. They are defined as 
the zero loci of the coefficient functions $f(z)$ and $g(z)$ 
of a Weierstrass equation. They are thought of as 
complex co-dimension one objects 
%, which we call 
%``elliptic point planes'', 
and correspond 
to the two kinds of critical points 
of a dessin d'enfant of Grothendieck. 
%
%We consider new complex co-dimension one objects 
%in F-theory on an elliptic fibration over $\PPsmall^1$,
%or a further fibration thereof, 
%consisting of zero loci of the coefficient functions 
%$f$ and $g$ of the Weierstrass equation, which 
%we referred to as an ``$f\!\!=\!\!0$-locus plane'' 
%and a ``$g\!\!=\!\!0$-locus plane'', collectively as 
%``elliptic point planes''.
%%If there are some elliptic point planes, 
%
The $\PPsmall^1$ base is divided into several cell regions
bounded by some domain walls extending from 
these planes and D-branes, 
on which the imaginary part of the $J$-function vanishes. 
%Each cell region corresponds to a (half of a) 
%fundamental region in the preimage of the $J$-function.  
This amounts to drawing a dessin with a canonical triangulation. 
We show that the dessin provides a new way of keeping track 
of mutual non-localness among 7-branes without 
employing unphysical branch cuts or their base point.
With the dessin we can see that   
weak- and strong-coupling regions coexist  
and are located across an $S$-wall from each other.
%We will see that, in the orientifold limit, such a ``non-perturbative 
%region'' shrinks to infinitely small so that it cannot be seen from 
%even a short distance. 
We also present a simple method for 
computing a monodromy matrix for an arbitrary path 
by tracing the walls it goes through.

\end{abstract}

\preprint{KEK-TH-2069}
%\pacs{}
\date{October12, 2019}

\maketitle

\newpage
\section{Introduction}
The importance of F-theory \cite{Vafa,MV1,MV2} in modern particle physics model building cannot be too much emphasized.  
The $SU(5)$ GUT, which can naturally explain the apparently 
complicated assignment of hypercharges to quarks and leptons, is readily achieved in F-theory.  Another virtue of F-theory is that it can yield 
matter in the spinor representation of $SO(10)$, into which all the 
quarks and leptons of a single generation are successfully incorporated, 
and which cannot be achieved in pure D-brane models. 
These features are shared by $E_8\times E_8$ heterotic models, 
but F-theory models have an advantage in that they may evade the issue 
of the relation between the GUT and Planck scales in heterotic string theory 
first addressed in \cite{Witten96}. Also, the Yukawa couplings 
perturbatively forbidden in D-brane models 
\cite{Yukawaforbidden1,Yukawaforbidden2} 
can be successfully generated in F-theory. 

Almost ten years after the first development in F-theory, 
there was much progress in the studies of local models of 
F-theory (See \cite{DonagiWijnholt,BHV,BHV2,
DonagiWijnholt2,HKTW,DWHiggsBundles,
localmodel1,localmodel2,localmodel3,localmodel4} 
for an incomplete list.).
In this class of theories, one basically considers a supersymmetric 
gauge theory\footnote{More precisely, the compact part of 
the theory is ``twisted'' so that the Casimirs of 
the gauge fields correctly transform as sections of 
Looijenga's weighted projective space bundle \cite{FMW}.} 
on a stack of 7-branes in F-theory,
whose coalescence is supposed to give rise to a gauge 
symmetry depending on the fiber type in the Kodaira 
classification. In particular, if the fiber type is either $IV^*$, 
$III^*$ or $II^*$, the gauge symmetry will be $E_6$, $E_7$ or 
$E_8$, respectively, and then the brane was called 
an exceptional brane \cite{BHV}. \footnote{More recently, 
after the LHC run in particular, 
global F-theory models have been attracting much interest. 
For recent works 
on global F-theory models, see e.g. \cite{globalmodel1,
AndreasCurio,
globalmodel2,
Collinucci,
BlumenhagenGrimmJurkeWeigand,
MarsanoSaulinaSchafer-NamekiThree-Generation,
BlumenhagenGrimmJurkeWeigand2,
MarsanoSaulinaSchaferNamekiU(1)PQ,
GrimmKrauseWeigand,
CveticGarcia-EtxebarriaHalverson,
ChenKnappKreuzerMayrhofer,
ChenChungE8point,
GrimmWeigand,
KnappKreuzerMayrhoferWalliser,
DolanMarsanoSaulinaSchaferNameki,
MarsanoSchaferNameki,
GrimmKerstanPaltiWeigand,
globalmodel3,globalmodel4,globalmodel5,globalmodel6,globalmodel7,
globalmodel8,globalmodel9,
CveticKleversPiragua,
BorchmannMayrhoferPaltiWeigand,
globalmodel10,globalmodel11,globalmodel12,
globalmodel13,globalmodel14,globalmodel15,globalmodel16}.
 }

The 
% v4 dessin %
%singularlity 
fiber
% v4 dessin %
type of such a codimension-one singularity 
can be labeled by the (conjugacy class of the) 
$SL(2,\ZZ)$ monodromy around the 
% v4 dessin %
%singularity.
fiber.
% v4 dessin %
It was shown that all the types of Kodaira 
% v4 dessin %
%singularities 
fibers
% v4 dessin %
can be represented by some product of monodromies of 
a basic set of 7-branes: {\bf A}=D-brane, 
{\bf B}=$(1,1)$-brane and {\bf C}=(1,$-1$)-brane \cite{GHZ,DZ,DHIZ},
as shown in Table in Appendix \footnote{In this paper,
we identify these 7-branes as the monodromy matrices 
$M_{p,q}$ defined in \cite{GHZ} with the sign of $q$ reversed 
(as we have adopted Schwarz's convention for the tension %(\ref{tension}
\cite{Schwarz9508143}), 
which are the 
{\em inverse} of $K_{[p,q]}$ in \cite{DZ,DHIZ}; 
this is consistent as the 
orderings of the branes and $K_{[p,q]}$'s are 
in reverse to each other.}.
The relation between the resolution of the 
% v4 dessin %
%Kodaira 
% v4 dessin %
singularity and the gauge symmetry on a coalescence of 
7-branes has been clearly explained by using 
string junctions. String junctions are also useful to 
describe chiral matter \cite{tani}, 
non-simply-laced Lie algebras \cite{BonoraSavelli},
the Mordell-Weil 
lattice of a rational elliptic surface \cite{FYY} 
and deformations of algebraic varieties \cite{GHS,GHS2}.

From the table one can see that the 
% v4 dessin %
%singularities 
singular fibers 
% v4 dessin %
of the exceptional type consist of a {\bf B}-brane 
and two {\bf C}-branes in addition to the ordinary 
D($=${\bf A})-branes. Thus, in this algebraic approach, 
the exceptional branes are seen to emerge due to the coalescence 
of these {\bf B}- and {\bf C}-branes which are distinct 
from D-branes. From a geometrical point of view, however, 
these branes are just the zero loci of the discriminant of 
a Weierstrass equation
% v4 dessin %
%\footnote{if the fibration has a section;  
%for models without sections see e.g. \cite{withoutsection} 
%and references therein.},
% v4 dessin % 
and there are no a priori 
differences from each other; they all are locally D-branes. 

% v4 dessin % 
%What makes an ordinary D-brane look like  
%a particular kind of $(p,q)$-brane is the ``{\em elliptic point 
%planes}'' %
%that we discuss in this paper. 
% v4 dessin % 

In this paper, we consider the special roles of the zero loci of 
the Weierstrass invariants $g_2(\tau(z))$, $g_3(\tau(z))$  
in F-theory on an elliptic fibration over $\PP^1$,  
or a further fibration thereof. They are defined as 
the zero loci of the coefficient functions $f(z)$ and $g(z)$ 
of a Weierstrass equation. They are thought of as 
complex co-dimension one objects, and we call them  
``{\em elliptic point planes}''.

% v4 dessin % 
In fact, mathematically, our construction amounts to 
drawing a  ``dessin d'enfant'' of Grothendieck 
on the $\PP^1$ base with a 
canonical triangulation\footnote{We thank the 
anonymous referee of Physical Review D for informing us 
of this fact.
}. 
We show that this drawing 
provides a new way of keeping track 
of mutual non-localness among 7-branes in place of 
the conventional {\bf A}{\bf B}{\bf C} 7-brane description. 
In our approach, all the discriminant loci are treated 
democratically, and 
with this ``dessin''  we can see that   
weak- and strong-coupling regions coexist  
and are located across an $S$-wall from each other. 
We also present a simple method for 
computing a monodromy matrix for an arbitrary path 
by tracing the walls it goes through.
The method for studying monodromies 
by tracing the contours on the $J(\tau)$-plane 
was developed long time ago by Tani \cite{tani}.

This paper is organized as follows:
%%%%%
In section 2, we introduce 
the basic setup of this paper, 
including the motivations and definitions of the elliptic point planes, 
the domain walls extended from 
them, and the cell region decomposition 
of the $\PP^1$ base of the elliptic fibration. The various 
definitions of the new notions and objects are summarized as 
a mini-glossary at the end of this section. 
% v4 dessin % 
In section 3, we briefly explain what is a ``dessin d'enfant'' 
and the relation to our present construction.
% v4 dessin % 
In section 4, 
we discuss the basic properties of the two kinds of elliptic point 
planes, the % v4 dessin % 
%$f$-locus planes
$f$-plane
% v4 dessin % 
 and the % v4 dessin % 
%$g$-locus planes.
$g$-plane. 
% v4 dessin % 
% v4 dessin % 
%In section 4, we explain how a {\bf C}(=$(1,-1)$)-7-brane 
%arises in the presence of the elliptic point planes, and 
%also study the substructure of an orientifold plane. 
% v4 dessin % 
In section 5, we present a new method 
for computing the monodromy by drawing the % v4 dessin % 
%wall chart
dessin.
% v4 dessin % 
%
%In section 6, we discuss how a string junction is described 
% in the present framework 
%as strings lightest in respective cell regions and 
%jointed near an % v4 dessin % 
%%$f$-locus plane
%$f$-plane 
%% v4 dessin % 
%.
%%%%% v3 %%%%%
%Section 7
%is devoted to the studies of all the types of the Kodaira 
%singularity achieved by D-branes and elliptic point planes. 
%Finally, in section 8, 
In the final section 
we conclude with a summary 
of our findings. Appendix A contains a table of 
% v4 dessin % 
%singularity 
fiber 
% v4 dessin % 
types 
of the Kodaira classification. 
The plots presented in this paper have been generated 
with the aid of Mathematica.    

%%%%%

\section{What is an elliptic point plane?} 
Consider a Weierstrass equation
\begin{eqnarray}
y^2=x^3+f x + g, \label{Weierstrasseq}
\end{eqnarray}
where $y$, $x$, $f$ and $g$ are sections of 
an ${\cal O}(3)$, an ${\cal O}(2)$, an ${\cal O}(4)$ and an ${\cal O}(6)$ 
bundle over the base $\PP^1$. This is a rational elliptic surface, 
which we regard as one of the two rational elliptic surfaces arising in the  
stable degeneration limit of a K3 surface. 
It may also be thought of as the total space of 
a Seiberg-Witten curve (with the ``$u$''-plane being the base)
of an ${\cal N}=2$ $SU(2)$ gauge theory or an E-string theory.
In an affine patch of $\PP^1$ with 
the coordinate $z$, the coefficient functions 
$f(z)$ and $g(z)$ are a 4th and a 6th order polynomial in $z$.
%%%%% v3 %%%%%
\footnote{Although we introduce and define various notions 
in this simple setup, 
most of them %(except the discussion in section 6) 
can be 
generalized to a lower-dimensional F-theory compactification 
on a higher-dimensional elliptic 
Calabi-Yau, whose base ${\cal W}$ 
is a $\PPsub^1$ fibration over some base manifold ${\cal B}$, 
by simply taking 
$y$, $x$, $f$ and $g$ to be sections of
$K^{-3}_{\cal W}$, 
$K^{-2}_{\cal W}$, 
$K^{-4}_{\cal W}$ and 
$K^{-6}_{\cal W}$, respectively, 
where $K_{\cal W}$ 
is the canonical class
of ${\cal W}$.  The equation (\ref{Weierstrasseq}) then
describes a K3 fibered Calabi-Yau over ${\cal B}$. 
A configuration of the elliptic point planes, 
D-branes and various walls are then 
a ``snapshot'' of a $\PPsmall^1$ fiber 
over some point on ${\cal B}$ with fixed coordinates. 
%The contents in section 6 strictly 
%apply only to a K3 compactification.
}

 As is well known, 
the modulus $\tau$ of the elliptic fiber of (\ref{Weierstrasseq}) is 
given by the implicit function: 
 \beqa
 J(\tau)&=&\frac{4f^3}{4f^3+27g^2},
 \eeqa
 where $J$ is the elliptic modular function.
The denominator of the right hand side 
\beqa
\Delta&\equiv&4f^3+27g^2
\eeqa
is called the discriminant. Near its zero locus $z=z_i$, 
$\mbox{Im}\tau$ goes to $\infty$ (if one has chosen 
the ``standard'' fundamental region) for generic 
(that is, nonzero) $f$ and $g$. Examining the behavior of   
$J(\tau)$ around $\infty$, we find 
\beqa
\tau(z)&=&\frac1{2\pi i}\log(z-z_i)\left(
\text{const.}+O(z-z_i)\right),
\eeqa
which implies the existence of a D7-brane at each discriminant locus.
%%%%% v3 %%%%%
\footnote{Thus, henceforth in this paper, we refer 
to a locus of the discriminant as (a locus of) a ``D-brane''.
As we will see, however,  
the monodromy around it is not always $T$ (\ref{T}) 
for a general choice of the reference point, due to the presence of the 
elliptic point planes.\label{footnoteDbrane}}
%%%%% v3 %%%%%

On the other hand, since a locus of 
% v4 dessin % 
%$f(z)$ or $g(z)$ 
$f(z)=0$ or $g(z)=0$ 
% v4 dessin % 
alone does not mean $\Delta=0$, it is not a D-brane. 
However, if the loci of  
% v4 dessin % 
%$f(z)$ and $g(z)$ 
$f(z)=0$ and $g(z)=0$ 
% v4 dessin % 
are present together 
with a D-brane, they play a significant role in generating a
$(p,q)$-7-brane by acting $SL(2,\ZZ)$ conjugate transformations 
on  a D-brane or as components of an orientifold plane, as we show below.  
In this paper, we will collectively call the loci of  
% v4 dessin % 
%$f(z)$ and $g(z)$ 
$f(z)=0$ and $g(z)=0$ 
% v4 dessin % 
{\em ``elliptic point planes''}.\footnote{In the standard fundamental region of the 
modular group of a two-torus, there are two elliptic points 
$\tau=e^{\frac{2\pi i}3}$ and $i$. They are fixed points 
of actions of some elliptic elements of $SL(2,\ZZsmall)$, 
hence the name. } 

%, and also call a  
% manifold which 
% contain such objects a 
%{\em monodromifold}, 
%and study their properties.
%\footnote{
%This definitions of  a ``monodromifold'' and a ``elliptic point plane'' 
%are different from those 
%of a conventional ``orientifold'' and an ``orientifold plane''.
%In the latter, one first defines an orientifold by using an  
%action of some discrete group on the space and and the worldsheet, 
%and then defines an 
%orientifold  (fixed) plane as a set of fixed points of this action 
%of the discrete group. In contrast, the definitions we provide 
%here for the former proceed in the reverse direction;   
%we first define a ``elliptic point plane'' as an object analogous 
%to an orientifold plane by using the Weierstrass equation, and then 
%define a ``monodromifold'' as an ambient space that contains 
%such objects afterwards. Technically, the ``monodromifold'' 
%defined in this way is almost nothing but the compact space 
%in F-theory, that is, an elliptically fibered space whose modulus is equal 
%to the complex scalar of type IIB theory, though,
%just like an orientifold, a ``monodromifold'' so defined 
%should not regard the elliptic fiber as a part of the space. 
%In any case,  what is important is not the nomenclature, but is 
%the orientifold-plane-like physical object itself we study, 
%whose relevance has never been emphasized in the literature. 
%}
%%%%% v3 %%%%%
%.
%%%%% v3 %%%%%
%The first example of this kind of compact space can be found in 
%\cite{cosmicstring}.

Elliptic point planes consist of two types, the loci of 
% v4 dessin % 
%$f(z)$ and $g(z)$, 
$f(z)=0$ and $g(z)=0,$ 
% v4 dessin % 
which have different properties. In this paper, we call the locus of 
% v4 dessin % 
%$f(z)$
$f(z)=0$ 
% v4 dessin % 
an 
% v4 dessin % 
%{\em $f$-locus plane},
{\em $f\!\!=\!0$ locus plane},
or an {\em $f$-plane} for short, 
% v4 dessin % 
and that of 
% v4 dessin % 
%$g(z)$
$g(z)=0$  
% v4 dessin % 
% v4 dessin % 
%a {\em $g$-locus plane}.
a {\em $g\!=\!0$ locus plane}, 
or a {\em $g$-plane} for short.
% v4 dessin % 
\footnote{
Despite the name ``plane'', an elliptic point plane is no 
more a rigid object but a smooth submanifold 
%of the ambient monodromifold 
when the elliptic fibration over $\PPsmall^1$ is further 
fibered over another manifold, 
just like a D-brane.} 
%%%%% v3 %%%%%
 
At the location of an 
% v4 dessin % 
%$f$-locus plane,
$f$-plane,
% v4 dessin % 
the value of the $J$-function 
is
\beqa
J(\tau)&=&\frac{4f^3}{4f^3+27g^2}~=~0, 
\eeqa
which corresponds to $\tau=e^{\frac{2\pi i}3}$. On the other hand, 
at the position of a % v4 dessin % 
%$g$-locus plane
$g$-plane 
% v4 dessin % 
, 
\beqa
J(\tau)&=&\frac{4f^3}{4f^3+27g^2}~=~1,
\eeqa
so this implies $\tau=i$.
In their neighborhoods, $J(\tau)$ is expanded as 
\beqa
J(\tau)&=&
\frac1{3!}J'''(e^{\frac{2\pi i}3})(\tau -e^{\frac{2\pi i}3})^3+O\left((\tau -e^{\frac{2\pi i}3})^4\right),
%&\simeq&
%-26.4728 i (\tau -e^{\frac{2\pi i}3})^3+O\left((\tau -e^{\frac{2\pi i}3})^4\right),
\label{Jflocusexpansion}\\
J(\tau)&=&
1-\frac{12 K\left(\frac{1}{\sqrt{2}}\right)^4 }{\pi ^2}(\tau -i)^2+O\left((\tau
   -i)^3\right),
%   &\simeq&
%1-14.3678(\tau -i)^2+O\left((\tau
%   -i)^3\right)
   \label{Jglocusexpansion}
\eeqa
%%%%% v3 %%%%%
where $K(k)$ is the complete elliptic integral of the first kind
\beqa
K(k)&=&\int_0^{\frac\pi2}\frac{d\theta}{\sqrt{1-k^2\sin^2\theta}}.
\eeqa
Thus
% so
%%%%% v3 %%%%% 
%
$\tau=e^{\frac{2\pi i}3}$ is a triple zero of $J(\tau)$ and 
$\tau=i$ is a double zero of  $J(\tau)-1$.

Suppose that 
$z=0$ is a locus of $f=0$. Since
\beqa
J(\tau(z))&=&\frac{4 f(z)^3}{4 f(z)^3+27 g(z)^2},
\label{Jtauz}
\eeqa
$J(\tau(z))$ is $O(z^3)$ at $z=0$. So 
(\ref{Jflocusexpansion}) shows that 
$\tau-e^{\frac{2\pi i}3}$ is $O(z)$ there, implying that the monodromy 
is trivial around the locus of $f$. 
Similarly, if $z=0$ is a locus of $g=0$, 
$J(\tau(z))-1$ is now $O(z^2)$. Comparing this with
(\ref{Jglocusexpansion}), we see that  $\tau(z)-i$ is also $O(z)$,
and hence there is no monodromy around the locus of $g=0$, either.

\begin{figure}[t]%
\centerline{
\includegraphics[width=0.5\textwidth]{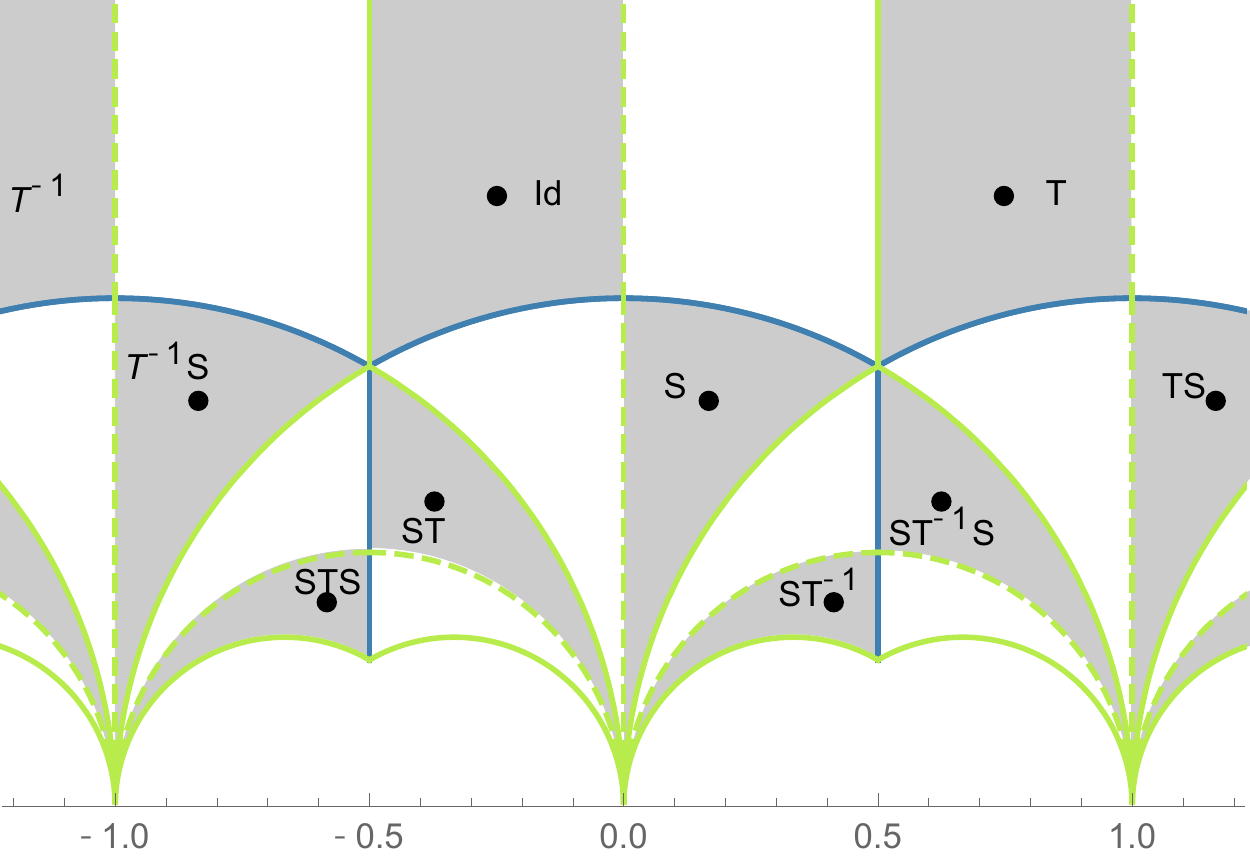}~~~
\includegraphics[width=0.45\textwidth]{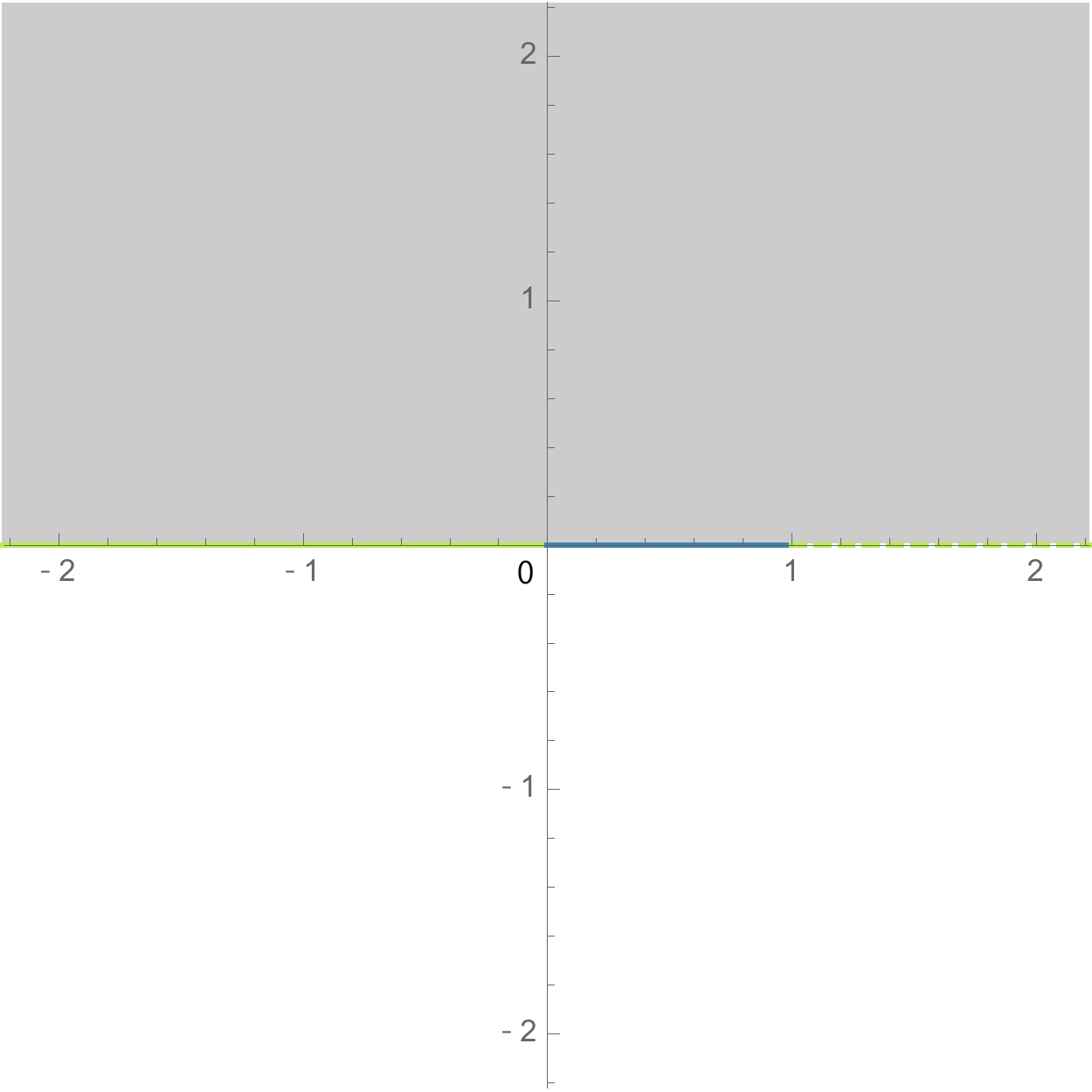}
}
\caption{\label{Teichmueller}Left: The 
% v4 dessin % Teichm\"uller space (= the
% v4 dessin %  
upper half plane
%)
and various fundamental regions. The shaded regions are 
the regions in which the imaginary part of the image of the 
$J$-function ${\rm Im}J(\tau)$ is positive. 
The symbol in each fundamental region (such as ${\rm Id}$, $T$, $S, \ldots$) 
is the group element of $SL(2,\ZZ)$ that maps the standard 
fundamental region to the fundamental region specified by the symbol.  
Right: The images of the $J$-function (= the whole complex plane). 
The green, blue and dashed green lines  
correspond to the respective boundary components of any
one half of  (the closure of) the fundamental regions.
}
\end{figure}

However, this is not the end of the story.  Figure \ref{Teichmueller} 
shows the various choices of 
fundamental regions of the modulus $\tau$ and the 
corresponding complex plane as its image mapped by the $J$-function.
From this we can see that if one goes around  
$\tau=e^{\frac{2\pi i}3}$ once on 
% v4 dessin % 
%the Teichm\"{u}ller space (=the upper half plane),
the upper half plane,
% v4 dessin % 
one goes through 
% v4 dessin % 
%{\em six} 
{\em three} 
% v4 dessin % 
different fundamental regions to get back 
to the original position.
Likewise if one goes around $\tau=i$,  
one undergoes {\em two} different fundamental regions. 
Thus an % v4 dessin % 
%$f$-locus plane
$f$-plane  
% v4 dessin % 
 is a complex codimension-one 
submanifold at which 
%
% v4 dessin % 
%six 
three 
% v4 dessin % 
%
different regions on the $z$-plane 
corresponding to different fundamental regions meet, 
while a % v4 dessin % 
%$g$-locus plane
$g$-plane 
% v4 dessin % 
 is similarly the place where two different 
regions meet. The regions on the $z$-plane 
corresponding to different fundamental regions are bounded 
by real codimension-one domain walls which consist of the zero 
loci of the imaginary part of the $J$-function. 

Furthermore, each region on the $z$-plane corresponding to 
a definite fundamental region is divided by a domain wall  
\beqa
\{\tau |~\mbox{Im}J(\tau)=0,~\mbox{Re}J(\tau)> 1 \}
\eeqa
(a dashed green line) into two regions 
 $\mbox{Im}J(\tau)>0$ and $\mbox{Im}J(\tau)<0$. 
 
\begin{figure}[h]%
\centerline{
\includegraphics[width=0.8\textwidth]{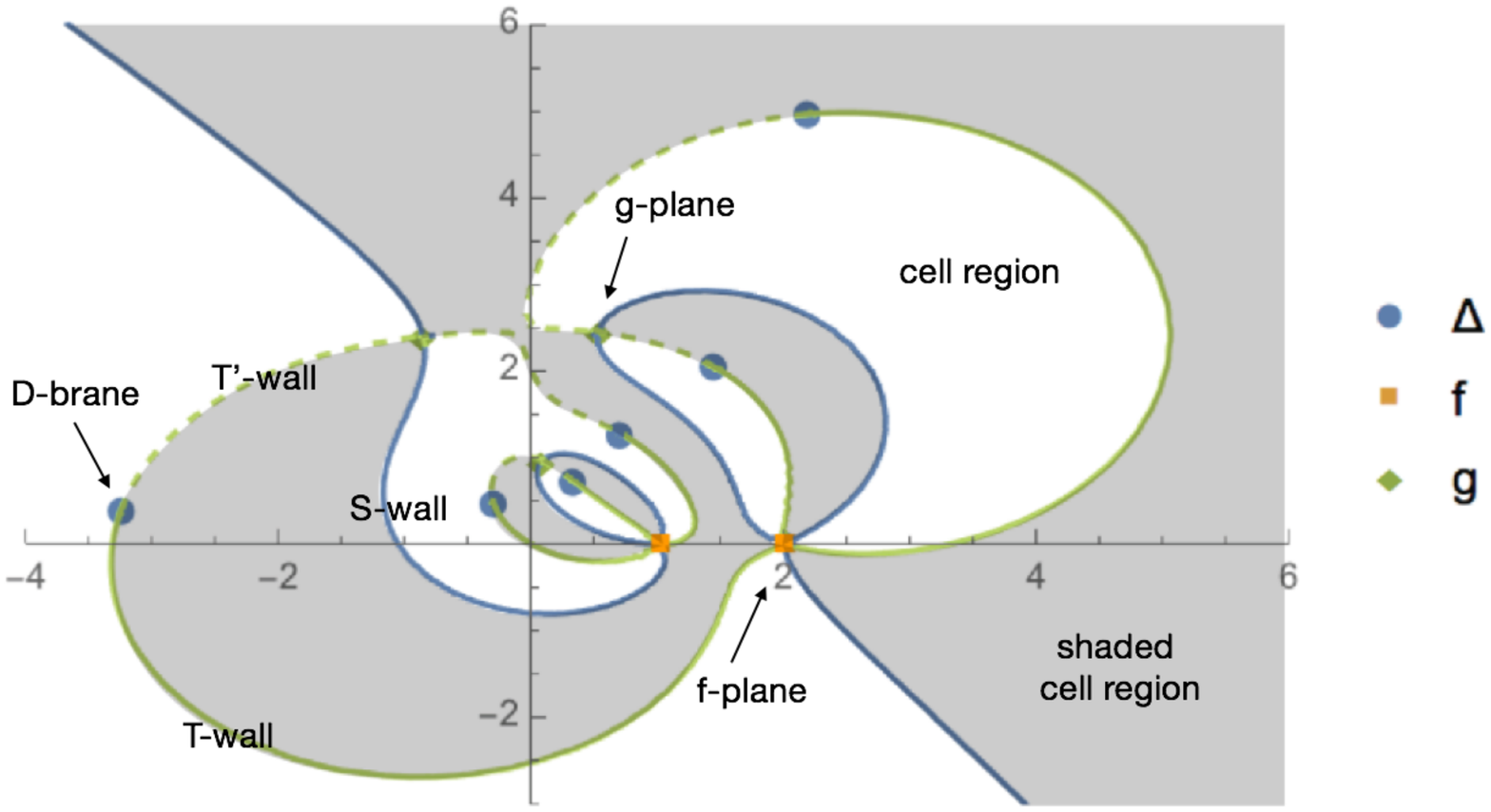}
}
\caption{\label{cell}
An example configuration of 
D-branes, elliptic point planes and the cell regions 
bounded by the domain walls extended from them.
D-branes are located at the loci of $\Delta=0$, while 
elliptic point planes are at the loci of $f=0$ and $g=0$.
In this example we can see two % v4 dessin % 
%$f$-locus planes
$f$-planes  
% v4 dessin % 
 at $z=1, 2$,
three % v4 dessin % 
%$g$-locus planes
$g$-planes 
% v4 dessin % 
 and six D-branes.
%%%%% v3 %%%%%
(This figure is depicted for the Weierstrass equation (\ref{Weierstrasseq})
for $f$ and $g$ (\ref{fgeq}) with $\epsilon=0.9$.)
%%%%% v3 %%%%%
}
\end{figure}

 On the other hand, a D-brane resides at a 
discriminant locus $\Delta=0$, from which 
two domain walls  $\{\tau |~\mbox{Im}J(\tau)=0,~\mbox{Re}J(\tau)< 0 \}$
(a green line)
and $\{\tau |~\mbox{Im}J(\tau)=0,~\mbox{Re}J(\tau)> 1 \}$ 
(a dashed green line)
extend out into 
the bulk $z$ space ($\PP^1$) (Fig.\ref{cell}).

Since the value of $J$ is $\infty$ 
at a discriminant locus for generic 
({\em i.e.} nonzero) values of $f$ and $g$, 
D-branes can never, by definition, touch nor pass through 
(a non-end point of)
the domain walls because $\mbox{Im}J(\tau)$ must vanish  
at the domain walls.

In this way, the $z$-space ($=\PP^1$) is divided into 
several ``{\em cell regions}'', which correspond to different 
fundamental regions in the preimage of the $J$-function,   
by the domain walls extended 
from the elliptic point planes (=% v4 dessin % 
%$f$-locus planes
$f$-planes 
% v4 dessin % 
 and % v4 dessin % 
%$g$-locus planes)
$g$-planes) 
% v4 dessin %  
and D-branes (Fig.\ref{cell}).
In particular, 
% v4 dessin % 
%$f$-locus planes
$f$-planes 
% v4 dessin % 
 and % v4 dessin % 
%$g$-locus planes
$g$-planes 
% v4 dessin % 
 extend
the domain walls 
\beqa
\{\tau |~\mbox{Im}J(\tau)=0,~0<\mbox{Re}J(\tau)< 1 \}
\label{S-wall}
\eeqa
(blue lines), and crossing through this wall 
implies that the type IIB coupling {\em locally gets S-dualized}  
(if starting from 
the standard choice of the fundamental region) 
(Fig.\ref{Teichmueller}).   
Then there is a difference in monodromies between when 
one goes around a D-brane within a single 
cell region bounded by some domain walls 
and when one first crosses through a 
domain wall, moves around a D-brane and then crosses back 
through the wall again to the original position; 
they are different by an $SL(2,\ZZ)$ conjugation.
This is what's happening in what has been called 
a ``{\bf B}-brane'' or a ``{\bf C}-brane'' in the discussions 
of string junctions.  
That is, while the monodromy matrix is necessarily 
\beqa
T&=&\left(
\begin{array}{cc}
1&1\\0&1
\end{array}
\right)
\label{T}
\eeqa
as long as the reference point is chosen to be 
in the standard fundamental region, 
a non-trivial (non-D-brane) $(p,q)$-brane arises 
if the monodromy is measured by going back and forth 
between regions corresponding to different fundamental 
regions in the preimage 
% v4 dessin % 
%Teichm\"uller space.  
upper-half plane.
% v4 dessin % 

% \usepackage[dvipdfmx]{graphicx}
%'Æ'µ'½'ç•\Ž¦'³'ꂽB

We would like to emphasize here that such a local 
$S$ transformation 
never takes place without these 
``elliptic point planes'' (=% v4 dessin % 
%$f$-locus planes
$f$-planes 
% v4 dessin % 
 and % v4 dessin % 
%$g$-locus planes
$g$-planes). 
% v4 dessin % 
%
 If it were not for elliptic point planes but there are only D-branes, 
the domain walls extended from them are only the ones 
\beqa
\{\tau |~\mbox{Im}J(\tau)=0,~\mbox{Re}J(\tau)< 0 \}
\label{T-wall}
\eeqa
(green lines)
and
\beqa
\{\tau |~\mbox{Im}J(\tau)=0,~\mbox{Re}J(\tau)> 1 \}
\label{T'-wall}
\eeqa
(dashed green lines).  So crossing through these walls only
leads to a $T$ transformation which commutes with 
the original monodromies of D-branes.
%\footnote{This is why we call 
%the loci of $f$ and $g$ ``elliptic point planes''.}

In the discussion below, we refer to 
the domain wall (\ref{T-wall}) (a green lines) as {\em $T$-wall} 
and the one (\ref{T'-wall}) (a dashed green line) as {\em $T'$-wall}, whereas 
we call the type of domain wall (\ref{S-wall}) (a blue line) {\em $S$-wall}.

To conclude this section we summarize the definitions of the new 
objects and notions introduced in this section as a mini-glossary.

\noindent
\underline{\em Mini-glossary}\\
{\bf\mbox{\boldmath $f$}-plane} A (complex) co-dimension-1 
object corresponding to a zero locus of $f(z)$ in the Weierstrass form
on the $z$-plane. Represented by a small square in the figures.
\\
{\bf\mbox{\boldmath $g$}-plane} A (complex) co-dimension-1 
object corresponding to a zero locus of $g(z)$ in the Weierstrass form
on the $z$-plane. Represented by a small $45^\circ$-rotated square
 in the figures.
\\
{\bf elliptic point plane} The collective name for % v4 dessin % 
%$f$-locus planes
$f$-planes 
% v4 dessin % 
and % v4 dessin % 
%$g$-locus planes
$g$-planes. 
% v4 dessin % 
\\
{\bf\mbox{\boldmath $T$}-wall} 
A (real) co-dimension-1 
object (domain wall) corresponding to a zero locus of ${\rm Im}J$
with ${\rm Re}J<0$, extending from a 
D-brane and a % v4 dessin % 
%$f$-locus plane
$f$-plane.
% v4 dessin % 
Represented by a green line.
\\
{\bf\mbox{\boldmath $T'$}-wall} 
A (real) co-dimension-1 
object (domain wall) corresponding to a zero locus of ${\rm Im}J$
with ${\rm Re}J>1$, extending from a 
D-brane and a % v4 dessin % 
%$g$-locus plane
$g$-plane. 
% v4 dessin % 
Represented by a dashed green line.
\\
{\bf\mbox{\boldmath $S$}-wall} 
A (real) co-dimension-1 
object (domain wall) corresponding to a zero locus of ${\rm Im}J$
with $0<{\rm Re}J<1$, extending from a 
% v4 dessin % 
%$f$-locus plane
$f$-plane 
% v4 dessin % 
 and a % v4 dessin % 
%$g$-locus plane
$g$-plane.
% v4 dessin % 
Represented by a blue line.
\\
{\bf cell region} A closed region on the $z$-plane ($\PP^1$ base 
of the elliptic fibration) bounded by the $T$-, $T'$- and $S$-walls. 
Each cell region corresponds to either half of the 
(closure of the) 
\footnote{Below we abuse terminology 
and refer to a ``fundamental region'' as 
one modulo points on its boundary.}
fundamental region with 
${\rm Im}J>0$ or ${\rm Im}J<0$ of the fiber modulus.
\\
{\bf shaded cell region} The cell region corresponding to the 
(closure of the) 
half fundamental region with ${\rm Im}J>0$ 
(Figure \ref{cell}).

% v4 dessin % 
\section{Relation to ``dessin d'enfant'' of Grothendieck}
In fact, 
the construction in the previous section is nothing but drawing a  
``dessin d'enfant'' of Grothendieck \cite{Grothendieck}, known in 
mathematics, on the $\PP^1$ base with a canonical triangulation.\footnote{
The contents of this section are triggered by a suggestion made by 
the anonymous referee of Phys.~Rev.~D.}
A dessin d'enfant, meaning a drawing of a child, is a graph consisting of 
some black points, white points and lines connecting these points, 
drawn according to a special rule.
To demonstrate the rule, let us consider, for example, 
a function \cite{LandoZvonkin}:
\beqa
F(x)=-\frac{(x-1)^3(x-9)}{64 x}=1-\frac{(x^2-6x-3)^2}{64 x}, 
\label{example}
\eeqa
where $x\in \PP^1$. $F$ is a map from $\PP^1$ to $\PP^1$. 
At almost everywhere on $\PP^1$,  $F$ is a homeomorphism, 
sending a small disk to another in a one-to-one way. However, 
$F$ maps a small disk centered at $x=1$ to one centered at $F=0$ 
in a three-to-one way. Similarly, $F$ is a two-to-one map from 
a small disk centered at $x=3\pm 2\sqrt{3}$ 
to one centered at $F=1$. The points $x=1,3\pm 2\sqrt{3}$ are said 
{\em critical points}, and the corresponding values of $F$ are 
said {\em critical values}. If the map from the neighborhood around a
critical point to another around the corresponding critical value is $k$-to-one, 
we say that the {\em ramification index} of the critical point is $k$.
 
 Now the rule to draw the dessin associated with (\ref{example}) is 
as follows: Place a black point at every preimage of $0$, and 
a while point at every preimage of $1$. Next draw lines at preimages 
of the line segment $[0,1]$. The result is shown in 
%the left panel of 
FIG.\ref{dessin_example}(a):
 \begin{figure}[h]%
\centerline{
\includegraphics[width=0.5\textwidth]{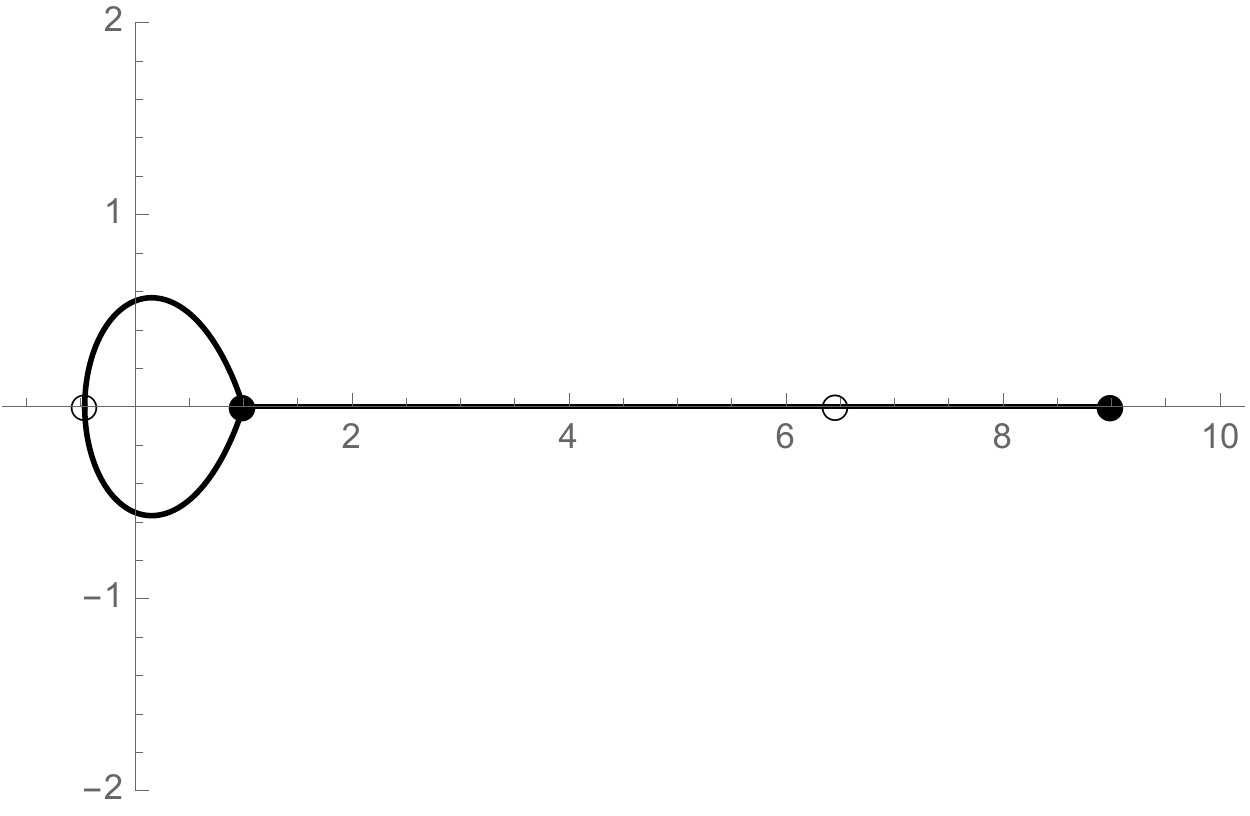}
\includegraphics[width=0.5\textwidth]{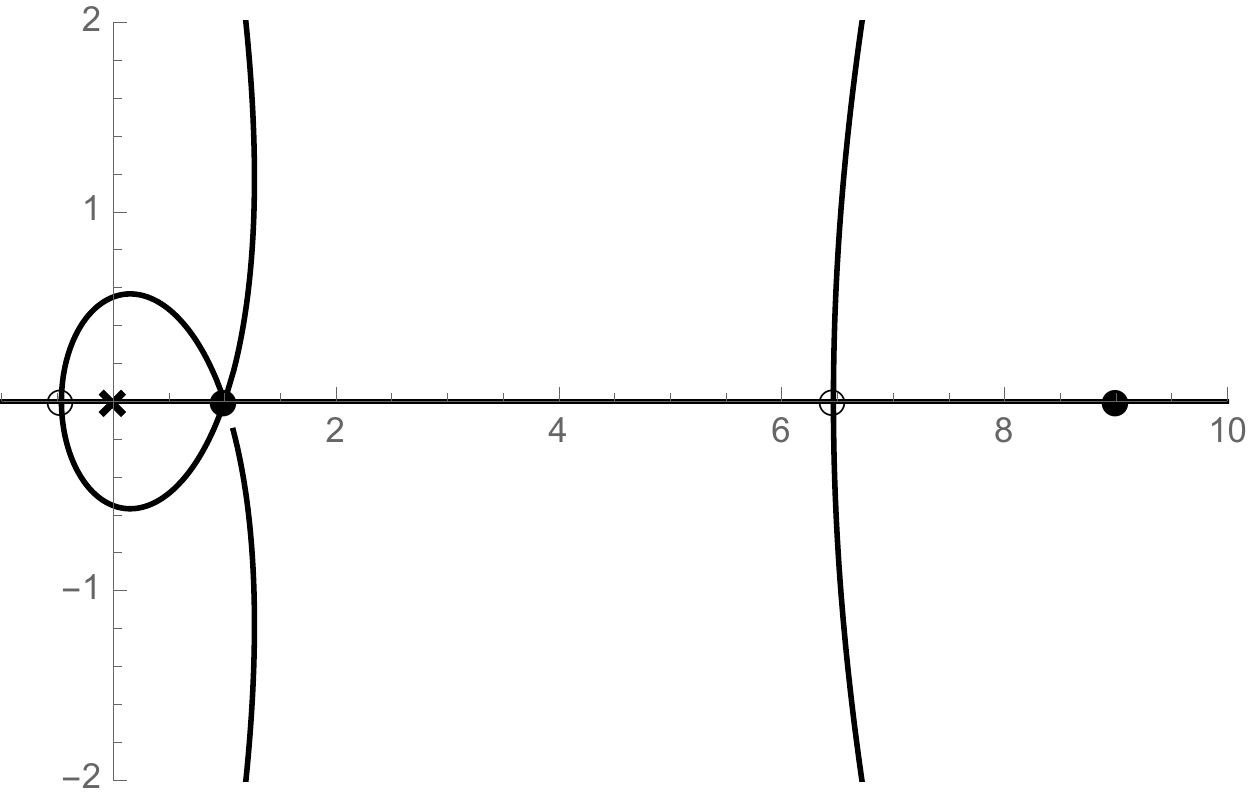}
}
\caption{\label{dessin_example}(a)(left panel): The dessin 
for (\ref{example}). (b)(right panel): The triangulated dessin. 
$\times$ represents an $\infty$ point. The extra lines 
have been drawn at the preimages of the segment $[-\infty,0]$ 
and $[1,\infty]$. The other $\infty$ point is not shown 
in this figure as it is infinitely far away.}
\end{figure}

The equation (\ref{example}) induces a branched covering over 
$\PP^1$. 
 Treating this graph as a combinatorial object, one can reproduce 
the information of the branched covering as follows: One first 
adds a point $\infty$ to each region of the dessin. One then 
connects each $\infty$ with lines to the black or 
white points as many times as they 
appear on the boundary of the region. This yields a triangulation 
of the dessin. Assigning either the upper- or the lower-half plane to 
each triangle depending on the ordering of $0$, $1$, $\infty$, 
and glueing these half planes together, 
one obtains a branched covering equivalent 
to the original one \cite{LandoZvonkin}.
 
In the present case, the equation (\ref{Jtauz}) 
defines a Belyi function, a holomorphic function 
whose critical values are only 
$0$, $1$ and $\infty$ and nothing else. %- of a special type. 
The black and white points in the dessin shown 
in FIG.\ref{dessin_example}(a) correspond to 
the $f$-planes and $g$-planes. The points  
$\infty$ added in the triangulation of the dessin are 
D-branes. 
The lines shown in FIG.\ref{dessin_example}(a) are the $S$-walls, 
while the lines connecting the $\infty$ points and 
the black or white points drawn in the triangulation are 
the $T$- and $T'$-walls.

What is special about (\ref{Jtauz}) is that it induces a 
local homeomorphism 
between the $\PP^1$ base and 
the upper-half plane. Indeed, as we saw 
in the previous section, the correspondence is one-to-one 
everywhere, even in the vicinity of the elliptic orbits 
$\tau=e^{\frac{2\pi i}3}$ 
and $i$. This is so because the $J=0$ ($f=0$) points are always  
critical points with ramification index three, and 
the $J=1$ ($g=0$) points are always with ramification index two.
In this paper, we treat the dessin not as just a 
combinatorial graph, but draw the $\infty$ points 
and the triangulating lines 
(the $T$- and $T'$-walls) also as preimages of the $J$-function, 
as shown in FIG.\ref{dessin_example}(b).
The special feature of (\ref{Jtauz}) then
allows us to use the (triangulated) 
dessin as a convenient tool to compute 
monodromies, as we see below.

% v4 dessin % 

\section{Basic properties of elliptic point planes}
\subsection{Basic properties of 
% v4 dessin %
%\mbox{\boldmath $f$}-locus planes
\mbox{\boldmath $f$}-planes
% v4 dessin %
}
As we defined in the previous sections, there are two kinds of 
elliptic point 
planes: % v4 dessin % 
%$f$-locus planes
$f$-planes 
% v4 dessin % 
 and % v4 dessin % 
%$g$-locus planes
$g$-planes.
% v4 dessin % 
In this section we describe the basic properties of % v4 dessin % 
%$f$-locus planes.
$f$-planes.
% v4 dessin % 

As the name indicates, 
% v4 dessin % 
%$f$-locus planes
$f$-planes 
% v4 dessin % 
 are the loci where the function $f$ vanishes.
As we saw in the previous section, these are the places where 
the $J$-function vanishes and $\tau$ becomes $e^{\frac{2\pi i}3}$ 
(or its $SL(2,\ZZ)$ equivalents).

As we saw in the previous section, the expansion of 
$J(\tau)$ near $\tau=e^{\frac{2\pi i}3}$ 
is given by (\ref{Jflocusexpansion}). 
If there is an % v4 dessin % 
%$f$-locus plane
$f$-plane 
% v4 dessin % 
 at $z=0$,
$f=0$ there, yielding 
\beqa
f(z)&=&f_{41}z+f_{42}z^2+\cdots,\\
g(z)&=&g_{60}+g_{61}z+g_{62}z^2+\cdots,
\eeqa
where $f_{4i}$, $g_{6j}$ are constants with indices 
running over $i=1,\ldots,8$ and $j=1,\ldots,12$ for a K3 surface
and $i=1,\ldots,4$ and $j=1,\ldots,6$ for a rational elliptic surface.
Since
\beqa
\frac{4 f^3}{4 f^3 + 27 g^2}&=&
\frac{4 f_{41}^3}{27 g_{60}^2} z^3 (1+ O(z)),
\eeqa 
$\tau(z)$ asymptotically approaches  
\beqa
\tau(z)&=&e^{\frac{2\pi i}3} + \frac{ 2 f_{41}}{(9 g_{60}^2 J'''(e^{\frac{2\pi i}3}))^{\frac13}} z
\label{tauflocus}
\eeqa
as $z\rightarrow  0$. Therefore, $\tau$ is regular 
near $z=0$, and hence an  
% v4 dessin % 
%$f$-locus plane
$f$-plane 
% v4 dessin % 
 does not carry D-brane charges. 

Parameterize a small circle around $z=0$ by 
 $z=\epsilon e^{i\theta}$ $(\epsilon>0)$, 
 then if one goes around along it once, 
 so does $\tau$ once around $e^{\frac{2\pi i}3}$ along 
 a small circle with a radius $\epsilon \left|
 \frac{ 2 f_{41}}{(9 g_{60}^2 J'''(e^{\frac{2\pi i}3}))^{\frac13}} 
\right|$. Thus, although the monodromy around an 
% v4 dessin % 
%$f$-locus plane
$f$-plane 
% v4 dessin % 
 is trivial, one passes through the boundary 
of the 
% v4 dessin % 
%fundamental 
half-fundamental 
% v4 dessin % 
region {\em six} times on the 
% v4 dessin %
%Teichm\"uller space 
upper-half plane
% v4 dessin %
as one goes once around an % v4 dessin % 
%$f$-locus plane
$f$-plane.
% v4 dessin % 
Since the neighborhoods of 
$z=0$ and $\tau=e^{\frac{2\pi i}3}$ are 
% v4 dessin % 
%isomorphic,
homeomorphic,
% v4 dessin % 
the neighborhood of $z=0$ around an % v4 dessin % 
%$f$-locus plane
$f$-plane 
% v4 dessin % 
is also divided into six cell regions corresponding to 
different % v4 dessin % 
%fundamental 
half-fundamental 
% v4 dessin %  
regions.
The six domain walls separating these cell regions consist of 
three $S$-walls (blue) with 
$(0<\mbox{Re}J(\tau)< 1)$ and three $T$-walls (green) 
$(\mbox{Re}J(\tau)< 0)$, which are extended alternately 
from the % v4 dessin % 
%$f$-locus plane
$f$-plane, 
% v4 dessin % 
forming a locally
 $\ZZ_3$-symmetric configuration.% (Fig?).

On the 
% v4 dessin %
%Teichm\"uller space, 
upper-half plane, 
% v4 dessin %
%{\em i.e.} the upper half plane, 
if one starts from the standard fundamental region and passes 
through preimages (of the $J$-function) 
of a $T$-wall (green) and an $S$-wall (blue) to go to 
the $SL(2,Z)$ equivalent point, then the $SL(2,\ZZ)$ transformation 
mapping the original point to the final point is $T^{-1} S$.
Further, if one crosses through preimages of 
a $T$-wall (green) and an $S$-wall (blue) 
again, the transformation to the final $SL(2,Z)$ equivalent point is 
$(T^{-1} S)^2=-ST\sim ST$ 
(as $PSL(2,\ZZ)$) .

Since 
\beqa
(T^{-1} S)^3&=& 1,
\eeqa
$T^{-1} S$ generates a $\ZZ_3$ group,
% v4 dessin %
which is the isotropy group of the elliptic point 
$\tau=e^{\frac{2 \pi i}3}$.
% v4 dessin %
It is easy to show that 
this $T^{-1} S$ transformation acts on the neighborhood of 
this point as a $\frac{2 \pi i}3$  rotation.
Therefore, the configuration of $\tau$ near an % v4 dessin % 
%$f$-locus plane
$f$-plane 
% v4 dessin % 
is locally 
invariant under the simultaneous actions of the spacial 
$\ZZ_3$ rotation and the $\ZZ_3$ $SL(2,\ZZ)$ transformation.
The metric near an % v4 dessin % 
%$f$-locus plane
$f$-plane 
% v4 dessin % 
 is locally $\ZZ_3$ invariant.

%%%%% v3 %%%%% 

%Therefore, the theory 
%is locally  $\ZZ_3$ invariant near an $f$-locus plane.
%If one considers equations of motion of other (say, spinor) fields 
%on this background, and if the solution satisfying the given 
%boundary conditions is unique, then the solution must also 
%be $\ZZ_3$ invariant. In this way, a $\ZZ_3$ invariance is 
%imposed locally in the vicinity of an $f$-locus plane. 
%Such a restriction on the fields is similar to ones encountered 
%in the case of orbifolds or orientifolds, but there are two differences: One is 
%that the restriction is imposed only locally, that is, on the configurations 
%infinitely close to the position of the $f$-locus plane. 
%Thus these restrictions give rise to some kind of 
%boundary conditions on solutions of 
%the equations of motion.  
%The other notable difference is that the ``twist'' 
%accompanied by a spacial $\ZZ_3$ rotation is an 
%$SL(2,\ZZ)$ $S$-dual transformation.

%%%%% v3 %%%%%

\subsection{Basic properties of 
% v4 dessin %
%\mbox{\boldmath $g$}-locus planes
\mbox{\boldmath $g$}-planes
% v4 dessin %
}
Likewise, the expansion of 
$J(\tau)$ around $\tau=i$ 
is given by (\ref{Jglocusexpansion}). 
Let a % v4 dessin % 
%$g$-locus plane
$g$-plane 
% v4 dessin % 
 be at $z=0$ this time. 
$f(z)$ and $g(z)$ are expanded as 
\beqa
f(z)&=&f_{40}+f_{41}z+f_{42}z^2+\cdots,\\
g(z)&=&g_{61}z+g_{62}z^2+\cdots.
\eeqa
Since 
 \beqa
\frac{4 f^3}{4 f^3 + 27 g^2}&=&
1-\frac{27 g_{61}^2}{4 f_{40}^3} z^2 (1+ O(z)),
\eeqa 
$\tau(z)$ approaches 
\beqa
\tau(z)&=& i + \frac{ 3i \pi^{\frac12} g_{61}}{4 K(\frac1{\sqrt{2}})^2 f_{40}^{\frac32}} z
\label{tauglocus}
\eeqa
as $z\rightarrow  0$. 
Thus $\tau$ is again regular near a % v4 dessin % 
%$g$-locus plane
$g$-plane, 
% v4 dessin % 
therefore a 
% v4 dessin % 
%$g$-locus plane
$g$-plane 
% v4 dessin % 
 does not have  
D-brane charges, either. The monodromy around a 
% v4 dessin % 
%$g$-locus plane
$g$-plane 
% v4 dessin % 
 is also trivial, although if one goes around it, one 
will be passing 
through the $S$-walls (blue lines) 
and the $T'$-walls (dashed green lines) alternately, 
twice for each.

Suppose that on the 
% v4 dessin %
%Teichm\"uller space 
upper-half plane
% v4 dessin %
one starts from an arbitrarily given point near 
$\tau=i$ in the standard fundamental region with 
$\mbox{Re}\tau<0$ and goes through the preimages 
of an $S$-wall and a $T'$-wall to 
reach the $SL(2,\ZZ)$-equivalent point. This move can be 
achieved by the
$SL(2,\ZZ)$ $S$ transformation.
This $S$ transformation acts on the neighborhood of $\tau=i$ 
as a $\ZZ_2$ rotation. The metric near a % v4 dessin % 
%$g$-locus plane
$g$-plane 
% v4 dessin % 
is also $SL(2,\ZZ)$ invariant.
% v4 dessin % 
Thus the vicinity of a % v4 dessin % 
%$g$-locus plane
$g$-plane 
% v4 dessin % 
 is 
invariant under the $\ZZ_2$ rotation associated with the $S$ 
transformation. 
\section{Simple method to compute the monodromy 
using the 
% v4 dessin % 
%wall chart
dessin
% v4 dessin % 
}
 Drawing the contours of the walls and the positions of 
the D-branes and elliptic point planes,  
we can have a figure of the complex plane divided into 
several cell regions such as 
FIG.\ref{cell}, %\ref{II}, \ref{I0startoI4} and \ref{I0startoI4shaded}, 
which we call a {\it % v4 dessin % 
%wall chart
dessin.}\footnote{This corresponds to a 
{\em triangulated} dessin in the sense of Grothendieck.}
% v4 dessin % 
 For a given 
Weierstrass equation, the % v4 dessin % 
%wall chart
dessin
% v4 dessin % 
 provides us 
with a very simple method to compute the monodromy 
matrices along an arbitrary path around branes 
on the complex plane 
(= an affine patch of the $\PP^1$ or the ``$u$-plane'' of 
a Seiberg-Witten curve).

\subsection{The method}
\begin{figure}[h]%
\vskip -20ex
\centerline{
\includegraphics[width=0.6\textwidth]{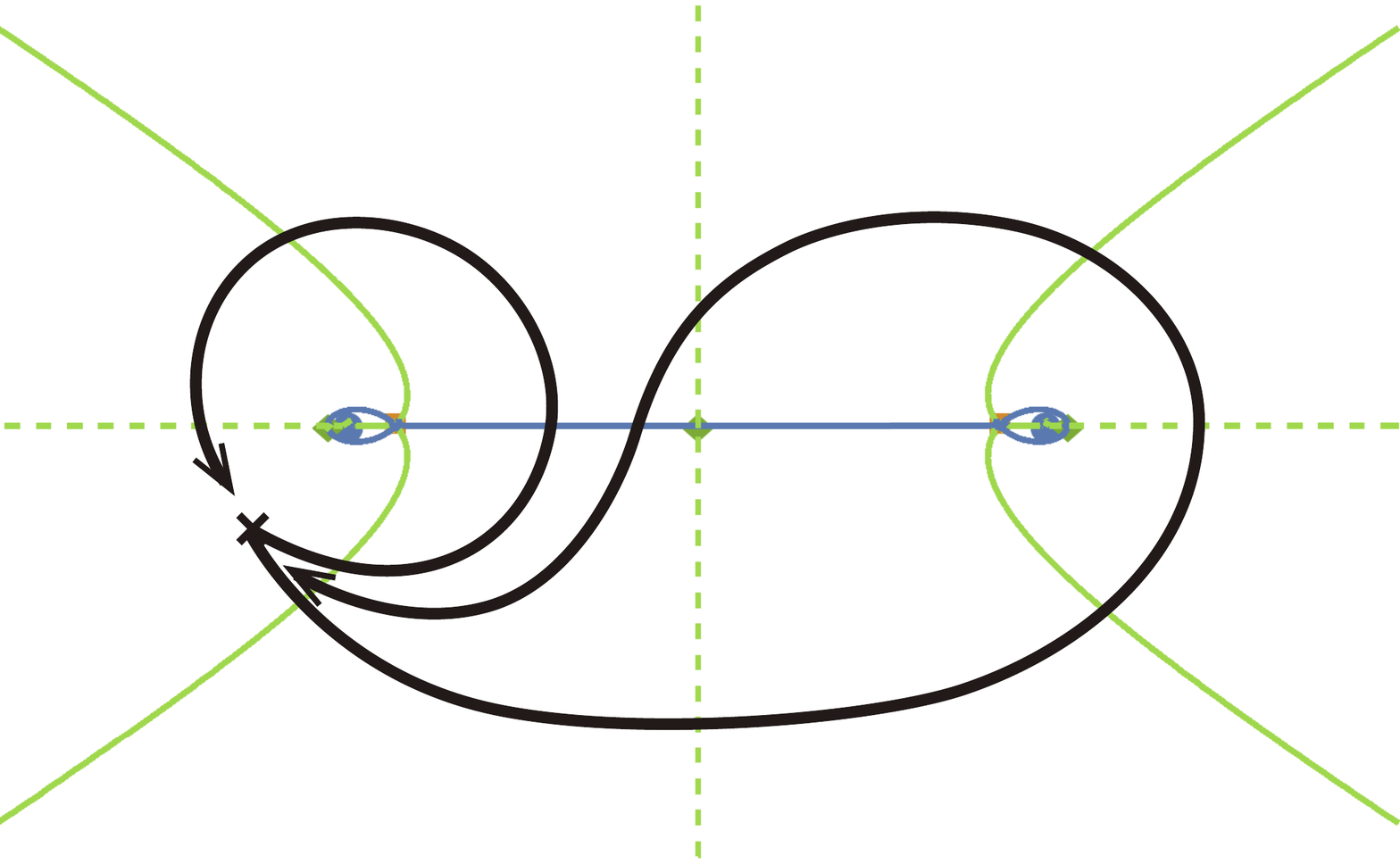}
}
\vskip -15ex
\centerline{
\includegraphics[width=0.6\textwidth]{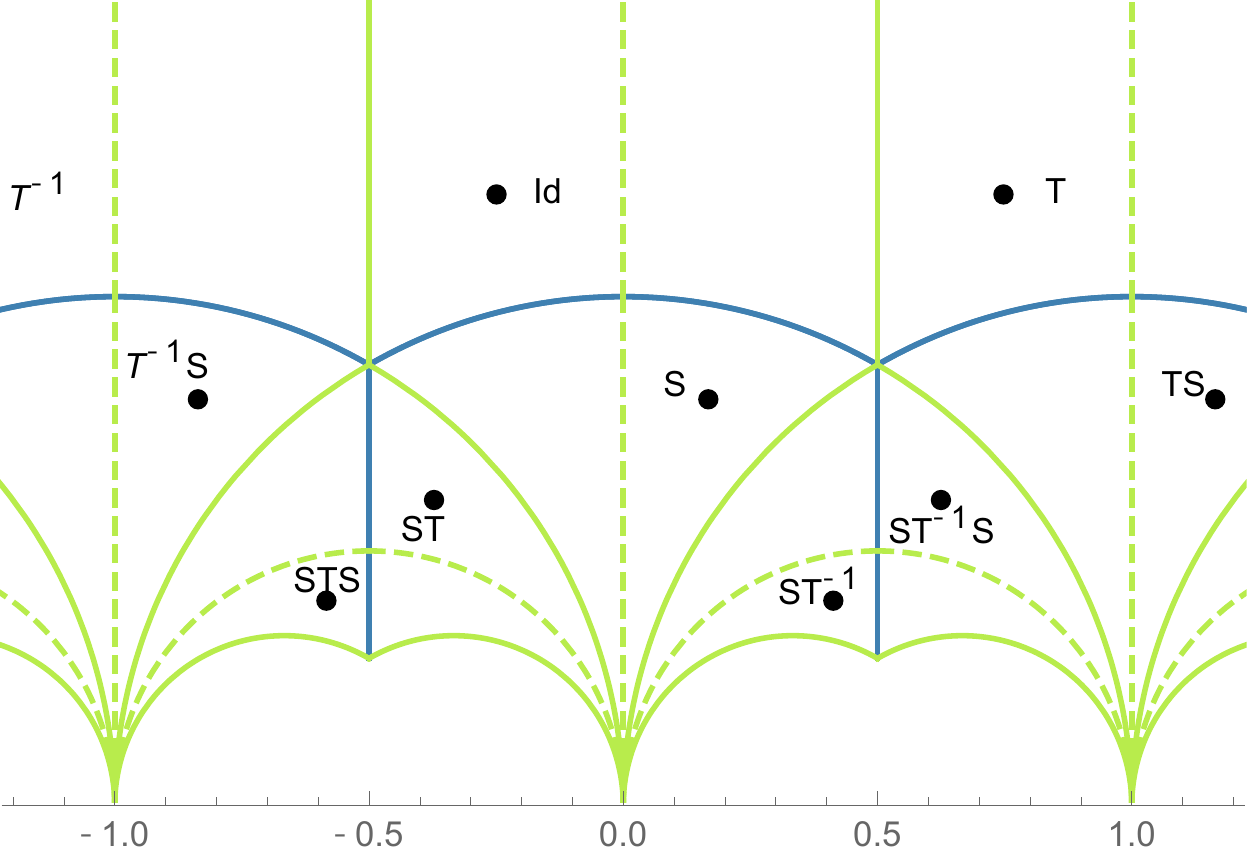}
}
%\vskip -5ex
\caption{\label{Nf=0SWcurve}
The upper panel: 
The % v4 dessin % 
%wall chart
dessin
% v4 dessin % 
 of 
$N_f=0$ SW curve ($f(u)=-\frac{1}3 u^2+1$, $g(u)=-\frac2{27} u^3 +\frac13 u$).
The lower panel: The crossed walls and the corresponding monodromies.
}
\end{figure}

To illustrate the method, let us consider the Seiberg-Witten curve 
of ${\cal N}=2$ pure ($N_f=0$) 
$SU(2)$ supersymmetric gauge theory \cite{SW}.
The equation is 
\beqa
y^2&=&x^3 -u x^2 + x.
\eeqa
Taking $u$ as the coordinate $z$, we obtain a Weierstrass equation 
with 
\beqa
f(u)=-\frac{1}3 u^2+1,~~~g(u)=-\frac2{27} u^3 +\frac13 u,
\eeqa
whose % v4 dessin % 
%wall chart
dessin
% v4 dessin % 
 is shown in the upper panel of Figure \ref{Nf=0SWcurve}.
Let us compute the monodromy around each discriminant locus. 
Choosing a starting point near the left locus
 (shown as a cross), the left path crosses the walls as
 \beqa
 \rightarrow {\bf G} \rightarrow {\bf B} \rightarrow {\bf G} \rightarrow {\bf dG} \rightarrow, 
 \label{leftpath}
 \eeqa
 where ${\bf G}$ denotes the $T$-wall, ${\bf B}$ the $S$-wall and ${\bf dG}$ the $T'$-wall.
 \footnote{
 ${\bf G}$, ${\bf B}$ and ${\bf dG}$ are respectively the first letters of 
 Green, Blue and dashed Green. We have avoided using $T$, $S$ or $T'$ here as 
 the monodromy matrices for the crossing do not coincide with the names 
 of the walls.
 }
% YamadaSWcurves.nb 
The monodromy matrices for various patterns of crossings are
\beqa
 \rightarrow {\bf dG} \rightarrow  {\bf G} \rightarrow &=&T, \n 
  \rightarrow {\bf G} \rightarrow  {\bf dG} \rightarrow &=&T^{-1}, \n 
  \rightarrow {\bf dG} \rightarrow  {\bf B} \rightarrow &=&  ~\rightarrow {\bf B} \rightarrow  {\bf dG} \rightarrow ~=~S, \n 
  \rightarrow~ {\bf B}~ \rightarrow  {\bf G} \rightarrow &=&ST, \n 
  \rightarrow ~{\bf G}~ \rightarrow  {\bf B} \rightarrow &=&T^{-1}S,
  \label{rule}
\eeqa
where the first wall of each row is the crossing 
from a shaded cell region (${\rm Im}J>0$) 
to an unshaded one (${\rm Im}J<0$), and the second is 
 from an unshaded to a shaded one. \footnote{Therefore, 
 these rules only apply when one computes a monodromy 
 for a path that starts from and ends in a {\em  shaded cell region} 
 (${\rm Im}J>0$). 
 The rules for computing a monodromy for a path from an {\em  un}shaded cell region (${\rm Im}J<0$) to another are similar but different:
 \beqa
 \rightarrow {\bf dG} \rightarrow  {\bf G} \rightarrow &=&T^{-1}, \n 
  \rightarrow {\bf G} \rightarrow  {\bf dG} \rightarrow &=&T, \n 
  \rightarrow {\bf dG} \rightarrow  {\bf B} \rightarrow &=&  ~\rightarrow {\bf B} \rightarrow  {\bf dG} \rightarrow ~=~S, \n 
  \rightarrow~ {\bf B}~ \rightarrow  {\bf G} \rightarrow &=&ST^{-1}, \n 
  \rightarrow ~{\bf G}~ \rightarrow  {\bf B} \rightarrow &=&TS.
  \label{ruleforunshaded}
\eeqa
 }
The monodromy matrices are defined as 
\beqa
T=
\left(
\begin{array}{cc}
1  & 1     \\
0  & 1   
\end{array}
\right),~~~S=
\left(
\begin{array}{cc}
0  & -1     \\
1  & 0   
\end{array}
\right)
\eeqa
as usual, where we say that the monodromy matrix is $\left(
\begin{array}{cc}
a  & b     \\
c  & d   
\end{array}
\right)$ if the modulus $\tau$ is changed to 
\beqa
\tau'~=~M\circ\tau~\equiv\frac{a\tau + b}{c\tau + d}. 
\eeqa
They are defined 
only in $PSL(2,\ZZ)$, {\it i.e.} up to a multiplication of $-1$.

By using the rule (\ref{rule}), we can immediately find the monodromy matrix 
for the path (\ref{leftpath}) as 
\beqa
T^{-1}S\cdot T^{-1}&=&T^{-1}ST^{-1}\n
&\sim&STS,
\eeqa
where $\sim$ denotes the equality in $PSL(2,\ZZ)$.

Similarly, the crossed walls for the right path are 
\beqa
 \rightarrow {\bf G} \rightarrow {\bf dG}  \rightarrow {\bf G} \rightarrow {\bf dG}
  \rightarrow {\bf G} \rightarrow {\bf dG} \rightarrow {\bf B} \rightarrow  {\bf G}  \rightarrow.
 \label{rightpath}
 \eeqa
Using rule (\ref{rule}) again, we find that the monodromy is 
\beqa
T^{-1}\cdot T^{-1}\cdot T^{-1}\cdot ST&=&T^{-3}ST.
\eeqa

A confusing but important point of the rule is that, in the first example,  
the monodromy matrix $T^{-1}$ which corresponds to 
the crossings  $ \rightarrow {\bf G} \rightarrow  {\bf dG} \rightarrow$
taking place {\em after} the crossings 
$ \rightarrow {\bf G} \rightarrow  {\bf B} \rightarrow$ 
is multiplied to $T^{-1}S$ from the {\em right}. This will be 
confusing because if 
$M=\left(
\begin{array}{cc}
a  & b     \\
c  & d   
\end{array}
\right)$, $M'=\left(
\begin{array}{cc}
a'  & b'     \\
c'  & d'   
\end{array}
\right)$ and 
 $\tau'=M\circ \tau$, %\frac{a\tau + b}{c\tau + d}$,  
 $\tau''=M'\circ \tau'$,  %\frac{a'\tau' + b'}{c'\tau' + d'}$,
 then the monodromy matrix $M''=\left(
\begin{array}{cc}
a''  & b''     \\
c''  & d''   
\end{array}
\right)$
 representing $\tau\mapsto \tau''=M''\circ\tau
 %\frac{a''\tau + b''}{c''\tau + d''}
 $ is
 given by 
 \beqa
 M''&= &M' M,
\label{M''=M'M}
 \eeqa
 in which $M'$ is multiplied from the {\em left}.

More generally, the following statement holds:
Let $\gamma$ be a path specified by 
the series of the walls 
\beqa
\gamma&:& \rightarrow {\bf W}_1 \rightarrow  
 {\bf W}_2\rightarrow \cdots  
 \rightarrow {\bf W}_k \rightarrow,
\eeqa
where ${\bf W}_{i}$ $(i=1,\ldots,k)$ are either of 
${\bf G}$, ${\bf B}$ or ${\bf dG}$, and 
let $M_\gamma$ denote the associated monodromy matrix 
of $\gamma$. $k$ is an even positive integer. 
(If it is odd,  
a shaded cell region is mapped to 
an unshaded cell region or vice versa, 
and the transformation cannot be an $SL(2,\ZZ)$ 
transformation).
Let $\gamma_1$, $\gamma_2$ be paths specified by 
the series of the walls crossed by them
\beqa
\gamma_1&:& \rightarrow {\bf W}^{(1)}_1 \rightarrow  
 {\bf W}^{(1)}_2\rightarrow \cdots  
 \rightarrow {\bf W}^{(1)}_{k_1} \rightarrow,\n
\gamma_2&:& \rightarrow {\bf W}^{(2)}_1 \rightarrow  
 {\bf W}^{(2)}_2\rightarrow \cdots  
 \rightarrow {\bf W}^{(2)}_{k_2} \rightarrow,
\eeqa
and let 
$\gamma_1+\!\!\!\!>\gamma_2$ be the jointed path
\beqa
\gamma_1+\!\!\!\!>\gamma_2&:& \rightarrow {\bf W}^{(1)}_1 \rightarrow  
 %{\bf W}^{(1)}_2\rightarrow 
 \cdots  
 \rightarrow {\bf W}^{(1)}_{k_1} \rightarrow
% \n
%&& 
{\bf W}^{(2)}_1 \rightarrow  
% {\bf W}^{(2)}_2\rightarrow 
\cdots  
 \rightarrow {\bf W}^{(2)}_{k_2} \rightarrow,
\eeqa
where we use the new symbol  $+\!\!\!\!>$ to denote 
the operation of jointing two paths.\footnote{We will not use 
the usual symbol for the addition ``$+$'' since this operation 
is noncommutative.}
Then
\\
\\
\noindent
{\bf Proposition.} 
\beqa
M_{\gamma_1+\!\!\!>\gamma_2}&=&M_{\gamma_1}M_{\gamma_2}.
\label{proposition}
\eeqa
{\it Remark.} 
As we noted above, the monodromy matrix corresponding to a 
later crossing comes to the {\it right}, unlike (\ref{M''=M'M}) in which the 
matrix for the later transformation is multiplied from the {\it left}.
\\
\\
\noindent
{\it Proof.} By induction with respect to the total number of crossed walls, 
it is enough to show the statement for the cases 
when $\gamma_2$ is any of the crossing patterns (\ref{rule}).
Suppose that $\gamma_1$ starts from a cell region $C_0$ and ends 
in another $C_1$, and that $\gamma_2$ goes from the 
cell region $C_1$ to another $C_2$,
where $\gamma_2$ is taken to be any of the crossing patterns (\ref{rule}), 
say, $\gamma_2=\rightarrow {\bf dG} \rightarrow  {\bf G} \rightarrow$ 
and $M_{\gamma_2}=T$.
Let $P_{\gamma_i}$ $(i=1,2)$ be 
the associated maps 
which send points in the cell region $C_{i-1}$ to those in the 
cell region 
$C_i$, respectively,
such that  the torus modulus over the point is 
$SL(2,\ZZ)$ equivalent. 
We say two points 
on $\PP^1$ are $SL(2,\ZZ)$ equivalent if the torus fiber moduli 
over them are $SL(2,\ZZ)$ equivalent. Using this terminology, 
we can say that $P_{\gamma_i}$ $(i=1,2)$  
are the maps which send the points in $C_{i-1}$ 
to their $SL(2,\ZZ)$ equivalent points in $C_i$,
respectively.
Since $\tau(z)$ is holomorphic 
in $z$ and $J(\tau)$ is holomorphic in $\tau$, 
the domain of the map $P_{\gamma_1}$ is not necessarily 
restricted to only $C_0$ but can be extended 
to outside $C_0$ as far as it is in a small neighborhood of $z_0$.

\begin{figure}[h]%
\vskip -5ex
\centerline{
\includegraphics[width=0.8\textwidth]{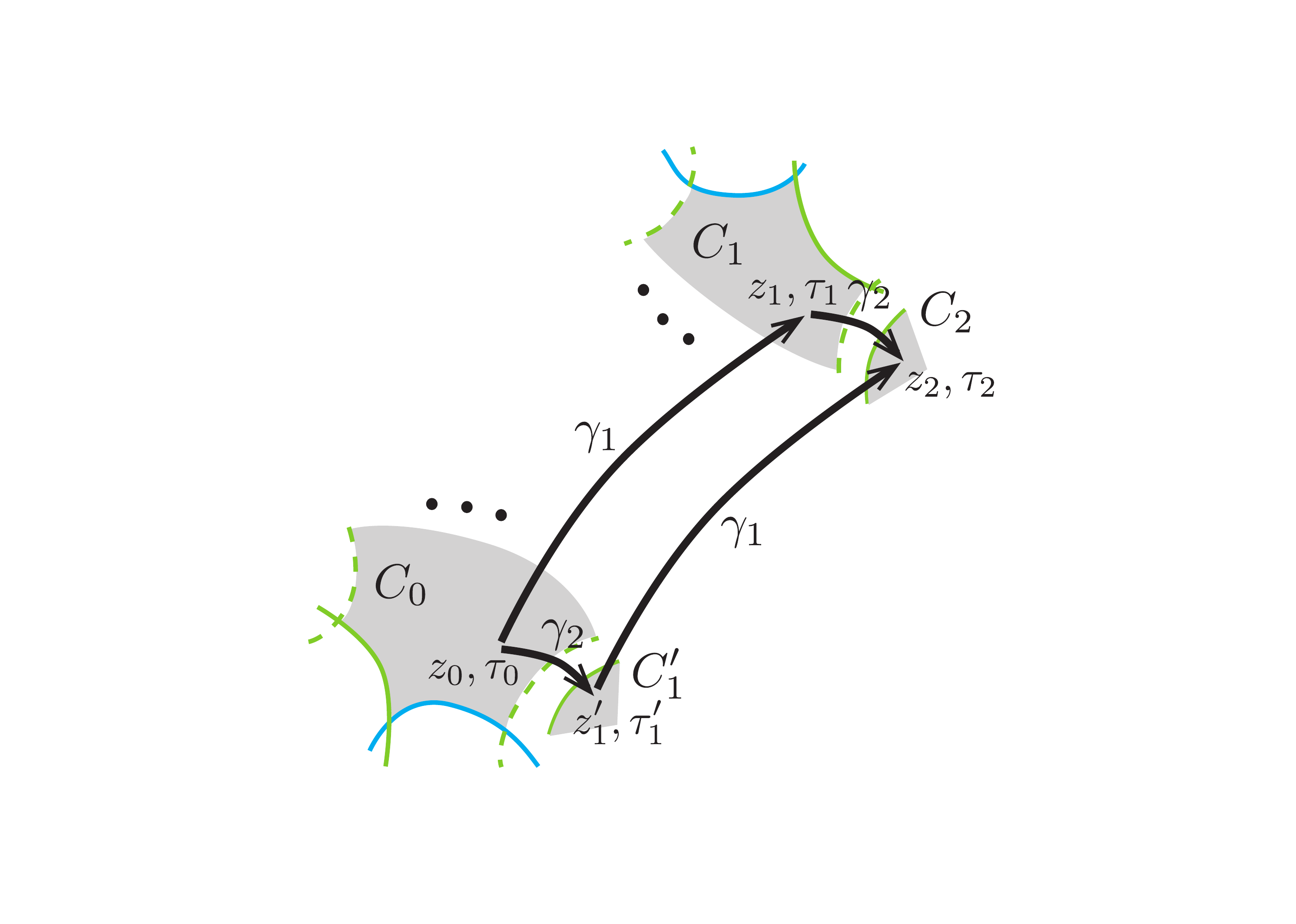}
}
\vskip -5ex
\caption{\label{proof}
Taking $\tau_0$ in the {\em standard} fundamental region, 
the transformation from $\tau_0$ to $\tau_1$ is given by
$\tau_1~=~M_{\gamma_1}\circ\tau_0$, but consecutive transformation 
from $\tau_1$ to $\tau_2$ is {\em not} $M_{\gamma_2}\circ\tau_1$,
as $\tau_1$ does not belong to the standard fundamental region 
in general. Rather, 
we have $\tau_2=M_{\gamma_1}\circ \tau'_1$ with
$\tau'_1=M_{\gamma_2}\circ \tau_0$ as 
$P_{\gamma_1}$ induces an isomorphism.
}
\end{figure}

Let $z_0$ be a point in $C_0$, and 
let $z_1=P_{\gamma_1}(z_0)\in C_1$, 
$z_2=P_{\gamma_2}(z_1)\in C_2$.
If we denote 
$\tau_i$ $(i=0,1,2)$ be the modulus of 
the torus fiber over $z_i$ $(i=0,1,2)$, they satisfy  
\beqa
J(\tau_i)&=&\frac{4f(z_i)^3}{4f(z_i)^3+27 g(z_i)^2},
\eeqa
where $\tau_1$ and $\tau_2$ are the values analytically continued 
from $\tau_0$ along the paths $\gamma_1$, 
and then $\gamma_2$.
Taking $\tau_0$ in the {\em standard} fundamental region, 
the transformation from $\tau_0$ to $\tau_1$ is given by
$\tau_1~=~M_{\gamma_1}\circ\tau_0$, but consecutive transformation 
from $\tau_1$ to $\tau_2$ is {\em not} $M_{\gamma_2}\circ\tau_1$,
as $\tau_1$ does not belong to the standard fundamental region 
in general. Rather, since $P_{\gamma_1}$ is locally an isomorphism
between a neighborhood around $z_0$ and that around $z_1$,
the final point  
$z_2$ can be written as the $P_{\gamma_1}$ image of $z'_1$,
where $z'_1$ is the $SL(2,\ZZ)$ equivalent point in the cell 
region  {\em reached along the path $\gamma_2$ first}
from $z_0$, if $z_2$ is close enough to 
$z_1$ (Figure \ref{proof}). If, on the other hand, $z_2$ is not close to 
$z_1$,  we can continuously deform the complex structure of 
the elliptic fibration so that $z_2$ may come close 
to $z_1$. Since this is a continuous deformation, the monodromy 
transformation matrix does not change, as the entries of 
the matrix take discrete values. Thus we may assume that $z_2$ 
is close to $z_1$.

Since $\tau_0$ is taken in the standard fundamental region, 
$\tau'_1$, the modulus of the torus fiber over $z'_1$, is 
given by
\beqa
\tau'_1&=&M_{\gamma_2}\circ \tau_0.
\eeqa   
Therefore, since $\tau_2=M_{\gamma_1}\circ \tau'_1$, 
we find
\beqa
\tau_2&=&M_{\gamma_1}\circ M_{\gamma_2}\circ \tau_0\n
&=&(M_{\gamma_1}M_{\gamma_2})\circ \tau_0,
\label{claim}
\eeqa
which is what the proposition claims. 

In deriving (\ref{claim}), we did not use the fact that 
$\gamma_2$ was assumed to be a particular pattern 
among (\ref{rule}), but the relation  (\ref{claim}) likewise holds
for other pattens. 
This completes the proof of the proposition.
\footnote{
In this proof, $\gamma_2$ is taken to be a path to the next adjacent 
cell region, whereas $\gamma_1$ is assume to be some 
long path leading to a faraway cell region. 
If $\gamma_1$ is also a path to another 
next adjacent cell region, it can be explicitly
checked that the proposition holds in this case as well. 
%However, if $\gamma_1$ is also short that leads to a next cell 
%region, one could also deform the configuration continuously 
%so that $z_2$ and $z'_1$ are also close. Then one could 
%change the roles of  $\gamma_1$ and $\gamma_2$ to 
%obtain a different (left-multiplication) rule, which would be inconsistent.
% What really happens is that, if $z_0$, $z_1$, $z_2$ and $z'_1$ 
% are all close, one cannot have both $P_{\gamma_1}$ and 
% $P_{\gamma_2}$ as isomorphisms unless $M_{\gamma_1}$ and 
% $M_{\gamma_2}$ commute. In any case,  
% it can be concretely checked that the proposition also 
% holds for such 
% cases.
}

\subsection{Example: Monodromies of $N_f=4$ $SU(2)$ Seiberg-Witten curves}
The proposition (\ref{proposition}) together with the rule (\ref{rule}) 
provides us with a very convenient method to compute the 
monodromy for an arbitrary Weierstrass model 
along an arbitrary path.

\begin{figure}[h]%
\vskip 0ex
\centerline{
\includegraphics[width=0.8\textwidth]{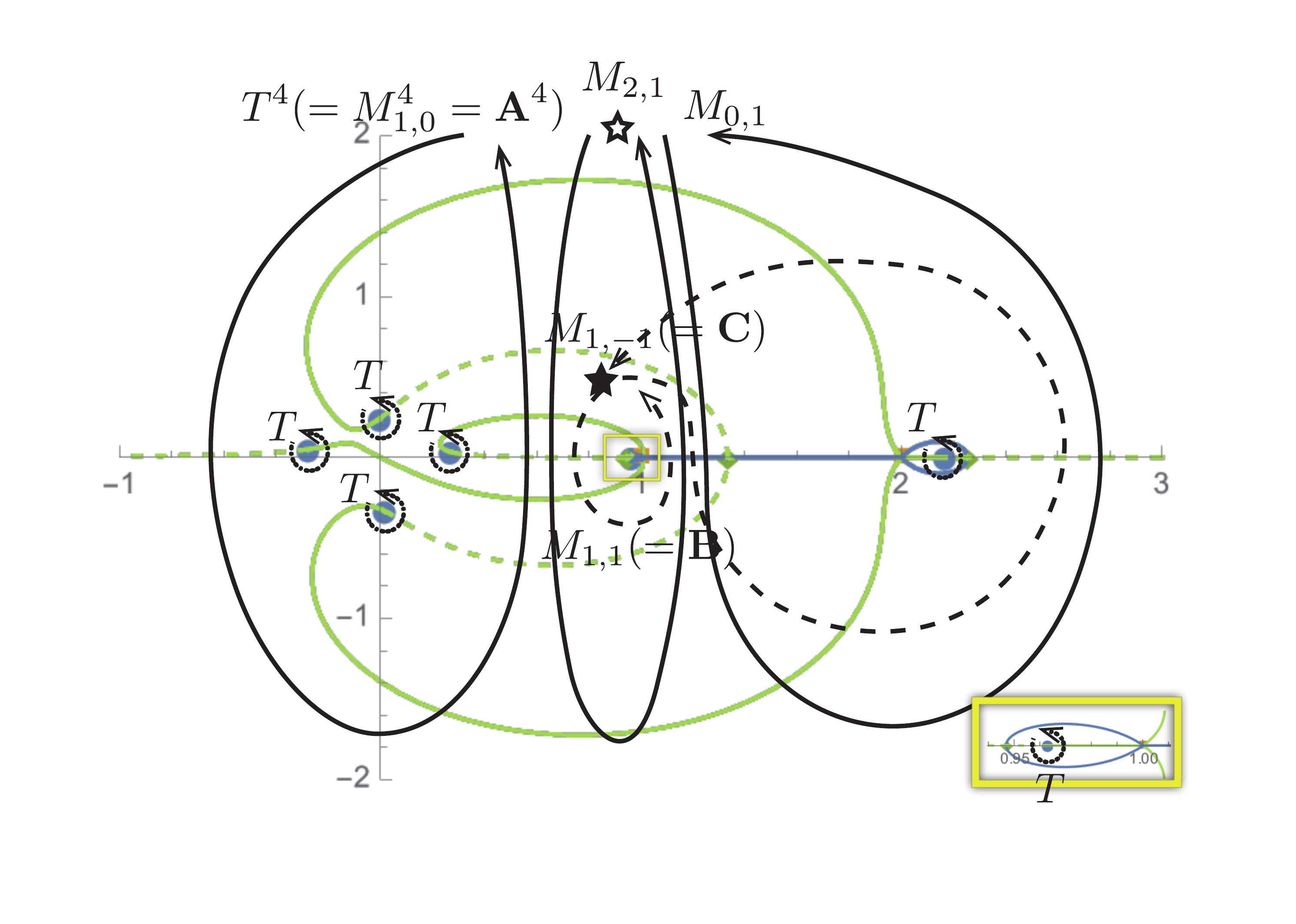}
}
\vskip 0ex
\caption{\label{Nf=4SWcurve}
Monodromies of $N_f=4$ $SU(2)$ Seiberg-Witten curve. 
It shows how the monodromies around the two D-branes 
on the right  (located at $z\approx 1$ and $\approx 2$) change 
depending on the choice of the reference point. If it is taken far enough 
(as marked as a white star), the monodromies along the black contours 
read 
$M_{2,1}$ and $M_{0,1}$. If the reference point is taken closer  
(as marked as a black star), then the monodromies along the dashed 
black contours are  $M_{1,1}(={\bf B})$ and $M_{1,-1}(={\bf C})$. 
If, on the other hand, the reference point is taken to be very close 
to the D-branes inside the cell regions surrounded by the $S$-walls, 
then the monodromies along the dotted contours are both $T$.}
\end{figure}

Figure \ref{Nf=4SWcurve} is a % v4 dessin % 
%wall chart
dessin
% v4 dessin % 
 of $N_f=4$ $SU(2)$ Seiberg-Witten curve
with some mass parameters.
The Weierstrass equation 
is (\ref{Weierstrasseq}) where
\beqa
f&=&(z-1)(z-2),\n 
g&=&\epsilon(z-i)(z-2i)(z-3i)\n
&&+(1-\epsilon)
\left(
-\frac{5}{16} i \sqrt{\frac{3}{2}} z^3+\frac{17 i z^2}{4 \sqrt{6}}-i \sqrt{6}
   z+\frac{4}{3} i \sqrt{\frac{2}{3}}
\right)
\label{fgeq}
\eeqa 
with $\epsilon=3\times 10^{-7}$. This choice of $g$ 
interpolates between the configuration in which all the $g$-locus 
planes are located on the imaginary axis at equal intervals 
($\epsilon=1$) and the one in which four of the six D-branes collide 
together at  $z=0$ to form a 
% v4 dessin %
%$I_4$ singularity 
$I_4$ singular fiber
% v4 dessin %
($\epsilon=0$), with 
the % v4 dessin % 
%$f$-locus planes
$f$-planes 
% v4 dessin % 
 fixed at $z=1,2$.
The figure is the configuration very close to the latter limit.

As is well known, the one-parameter (``$u$'') family of tori 
describe the moduli space of the gauge theory and can be 
compactified into a rational elliptic surface by taking 
the variables and coefficient functions to be sections of 
appropriate line bundles, where the $u$ parameter becomes 
the affine coordinate $z$ of the base $\PP^1$. 
Note, however, that the % v4 dessin % 
%wall chart
dessin
% v4 dessin % 
 %
can be drawn on this affine patch independently 
of the choices of the bundles; it only affects 
how many D-branes are at the infinity of $\PP^1$.  

This figure shows how the monodromies around the two D-branes 
on the right  (located at $z\approx 1$ and $\approx 2$) change 
depending on the choice of the reference point. If it is taken far enough 
(as marked as a white star), the monodromies along the black contours 
read $M_{2,1}$ and $M_{0,1}$. This means that, as we show later, 
a $(2,1)$ and a $(0,1)$ string become light near the respective 
D-branes, showing that the locations of the D-branes are the $(2,1)$ 
dyon and the monopole point on the moduli space of the gauge 
theory, which is well known.

 If the reference point is taken closer  
(as marked as a black star), then the monodromies along the dashed 
black contours are  $M_{1,1}(={\bf B})$ and $M_{1,-1}(={\bf C})$, 
which agrees with the {\bf ABC} brane description of the $I_0^*$ 
Kodaira 
% v4 dessin % 
% singularity.
singular fiber. 
 % v4 dessin % 

Finally, if the reference point is taken to be very close 
to the D-branes inside the cell regions surrounded by the $S$-walls, 
then the monodromies along the dotted contours are both $T$, 
showing that these branes look ordinary D-branes if they are observed 
from very close to them.

\subsection{\mbox{\boldmath $(p,q)$}-brane as an effective description
 % v4 dessin % 
 %of a cluster of a D-brane and elliptic point planes
 % v4 dessin % 
 }
Of course, it is well known that the monodromy changes depending 
the choice of the reference point. A monodromy matrix measured 
from some reference point gets $SL(2,\ZZ)$ conjugated if it is 
measured from another point.
What is new here that, by drawing a % v4 dessin % 
%wall chart
dessin,
% v4 dessin % 
we can precisely 
see how and from where the monodromy matrix changes and 
gets conjugated as we vary the position of the reference point.

For instance, we can see from Figure \ref{Nf=4SWcurve} 
that the monodromies 
around the two D-branes on the right are either $M_{2,1}$, $M_{0,1}$ 
or $M_{1,1}(={\bf B})$, $M_{1,-1}(={\bf C})$ for most choices of 
the reference point on the $z(\equiv u)$-plane, and they are recognized 
as ordinary ($M_{1,0}={\bf A})$ D-branes only when they are viewed 
from the points in the tiny regions surrounded by the $S$-walls. 
Thus we see that the effective description of the two branes 
as $(1,1)(={\bf B})$- and $(1,-1)(={\bf C})$-branes are good 
at the energy scale lower than the scale of the size of the 
small cell regions surrounded by the $S$-walls.

However, one can also set the mass parameters of the same 
gauge theory so that the % v4 dessin % 
%wall chart 999
dessin
% v4 dessin % 
 of the Seiberg-Witten curve 
looks as shown in Figure 
% v4 dessin % 
%\ref{anotherNf=4SWcurve}. 
%The Weierstrass 
%equation is given by the same $f$ and $g$ (\ref{fgeq}) 
%but $\epsilon=1$.
\ref{cell}.
%\begin{figure}[h]%
%\vskip 0ex
%\centerline{
%\includegraphics[width=0.6\textwidth]{epsilon1limit.pdf}
%}
%\vskip 0ex
%\caption{\label{anotherNf=4SWcurve}$N_f=4$ SW curve
%with another set of parameters.
%%
%}
%\end{figure}
% v4 dessin % 
%
In this case, the $S$-walls spread into wide areas of the 
$\PP^1$. There is not much difference among 
the six D-branes, and there is no obvious reason %distinction
to distinguish particular two as {\bf B} or {\bf C} 
from the other four D-branes.
% v4 dessin % 
%\footnote{though the D-brane connected 
%to the $f$-locus at $z=1$ 
%by a $T$-wall is already surrounded by $S$-walls;
%this eventually becomes ``{\bf B}''   
%when $\epsilon\approx 0$. 
%However, the one connected to the $f$-locus at $z=2$,
%which is eventually identified as ``{\bf C}'', 
%can be joined to the leftmost D-brane 
%by a path without crossing any $S$-walls. This happens 
%due to a recombination of walls, as we discuss in the next 
%section.}\\
% v4 dessin % 
\\

\noindent{\em Remark.} 
We have seen that a cluster of a D-brane and two 
elliptic point planes, in which the former is surrounded 
by the $S$-walls extended from the latter, may be effectively 
identified as a {\bf B}- or a {\bf C}-brane, if viewed from a
distance of the size of the cluster. Thus one might think that 
an ``exact'' $(p,q)$-brane 
(whose monodromy is $M_{p,q}$ along arbitrary small loop)
can be obtained by taking the $f$-
 and % v4 dessin % 
%$g$-locus planes
$g$-planes 
% v4 dessin % 
 on top of each other so that the
 size of the cell region the $S$-walls surround
becomes zero. This is not the case,  however, 
since if the $f$- and % v4 dessin % 
%$g$-locus planes
$g$-planes 
% v4 dessin % 
 collide, the order 
of the discriminant becomes two, implying that 
another D-brane also automatically comes on top of 
the D-brane, % v4 dessin % 
%$f$-locus plane
$f$-plane 
% v4 dessin % 
 and % v4 dessin % 
%$g$-locus plane
$g$-plane. 
% v4 dessin % 
Since it contains two D-branes, it cannot be identified  
as a {\em single} $(p,q)$-brane %(it is rather a Kodaira fiber of type II) 
in the {\bf ABC}-brane description.

\section{Conclusions}

The coexistence of D-branes and 
non-pure-D-7-branes is an essential feature of F-theory,  
as it 
enables us to achieve exceptional group gauge 
symmetries or matter in spinor representations
%%%%% v3 %%%%%
by allowing string junctions to appear as extra objects 
ending on more than two 
different types of 7-branes, 
in addition to the open strings which can only connect 
two ordinary D-branes.  
%%%%% v3 %%%%%
These 7-branes are conventionally described algebraically  
in terms of {\bf ABC} 7-branes. 
In this paper, noticing that all the discriminant loci are on equal footing 
and there is no a priori reason to distinguish one from the others, 
we have considered new 
%%%%% v3 %%%%%
complex co-dimension one
%%%%% v3 %%%%%
objects consisting of 
the zero loci of the coefficient functions $f$ and $g$ of the 
Weierstrass equation, which we referred to as an % v4 dessin % 
%$f$-locus plane
``$f$-plane'' 
% v4 dessin % 
and a % v4 dessin % 
%$g$-locus plane
``$g$-plane'', 
% v4 dessin % 
collectively as ``elliptic point planes''.
% v4 dessin % 
They are two kinds of critical points of 
a ``dessin d'enfant'' known in mathematics.
% v4 dessin % 

Although they do not carry D-brane charges, 
they play an essential role 
in achieving an exceptional gauge symmetry and/or 
a spinor representation  
by altering the monodromies around the branes. More precisely, 
if there are some elliptic point planes, the $z$-plane is divided 
into several cell regions, each of which corresponds to a (half of a) 
fundamental region in the preimage of the $J$-function. 
A cell region is bounded by several domain walls extending from 
these elliptic point planes and D-branes, on which the imaginary 
part of the $J$-function vanishes. In particular, the elliptic point 
planes 
extend a special kind of domain walls, which we call ``$S$-walls'', 
crossing through which implies that the type IIB complex 
string coupling is $S$-dualized. Consequently, on the $z$-plane 
coexist a theory in the perturbative regime and 
its nonperturbative $S$-dual simultaneously. %This is F-theory.
The monodromy around several 7-branes is thus not just a 
product of monodromy around each 7-brane any more,  
but they get $SL(2,\ZZ)$ conjugated due to the difference of 
the corresponding fundamental regions the base points 
%of the monodromies 
belong to. 
% v4 dessin % 
%This is the origin of the $(p,q)$-7-branes; 
%what makes exceptional branes exceptional is the elliptic point planes.
% v4 dessin % 

In this sense one may say that the nonperturbative properties of 
F-theory --- the realizations of exceptional group symmetry, 
matter in spinor representations, etc. --- are the consequence of the 
coexisting ``locally $S$-dualized regions'' bounded by the $S$-walls 
extended from the elliptic point planes.  
In the orientifold limit \cite{Senorientifold}, the D-branes and the 
 %
 % v4 dessin % 
% monodromifold 
 elliptic point 
 % v4 dessin %  
 planes gather to form a $I_0^*$ 
 %
 % v4 dessin % 
% singularity, 
singular fiber, 
 % v4 dessin % 
so that the $S$-walls 
extended from the elliptic point planes are contracted 
with each other and confined, so the $S$-walls 
are not seen from even a short distance.

 % v4 dessin % 
We hope this new way of presenting the non-localness 
among 7-branes will be useful for understanding of the 
structure of higher-codimension singularities with 
higher-rank enhancement such as discussed 
in \cite{MV1,MV2,BIKMSV,HKTW,
MorrisonTaylor,FtheoryFamilyUnification,MizoguchiTaniAnomaly}.

\section*{Acknowledgments}
We wish to thank the referee of Phys.~Rev.~D 
for suggesting the improvement of the manuscript 
by considering the mathematical concept of dessin d'enfant. 
We also thank Y.~Kimura and T.~Tani for valuable discussions. 
The work of S. M. is supported by 
Grant-in-Aid for Scientific Research (C) \#16K05337 
from The Ministry of Education, Culture, Sports, 
Science and Technology of Japan.

%\newpage
\section*{Appendix}
\begin{table}[ht]
\caption{\label{Kodaira}The Kodaira  
classification. %\cite{Kodaira,DHIZ}. 
ord$(f)$, ord$(g)$ and ord$(\Delta)$ denote the orders of zeros of 
$f$, $g$ and the discriminant $\Delta$ of the Weierstrass equation.
}
\centering
%\begin{ruledtabular}
\begin{tabular}{|c||c|c|c|c|c|c|}
\hline
Fiber type&
ord$(f)$&
ord$(g)$&
ord$(\Delta)$&
Singularity type&
7-brane configuration&
Brane type\\
\hline
$I_n$
&$0$
&$0$
&$n$
&$A_{n-1}$
&$\AA^n$
&$A_{n-1}$
\\
$II$
&$\geq 1$
&$1$
&$2$
&$A_0$
&$\CC\AA$
&$H_0$
\\
$III$
&$1$
&$\geq 2$
&$3$
&$A_1$
&$\CC\AA^2$
&$H_1$
\\
$IV$
&$\geq 2$
&$2$
&$4$
&$A_2$
&$\CC\AA^3$
&$H_2$
\\
$I_n^*$
&$\geq 2$
&$3$
&$6+n$
&$D_{n+4}$
&$\AA^{n+4}\BB\CC$
&$D_{n+4}$
\\
$I_n^*$
&$2$
&$\geq 3$
&$6+n$
&$D_{n+4}$
&$\AA^{n+4}\BB\CC$
&$D_{n+4}$
\\
$II^*$
&$\geq 4$
&$5$
&$10$
&$E_8$
&$\AA^7\BB\CC^2$
&$E_8$
\\
$III^*$
&$3$
&$\geq 5$
&$9$
&$E_7$
&$\AA^6\BB\CC^2$
&$E_7$
\\
$IV^*$
&$\geq 3$
&$4$
&$8$
&$E_6$
&$\AA^5\BB\CC^2$
&$E_6$
\\
\hline 
\end{tabular}
%\end{ruledtabular}
\end{table}

\end{document}